\documentclass[11pt]{article}

\usepackage[a4paper,hmarginratio=1:1,vmarginratio=2:3,totalwidth=15.1cm,totalheight=22.7cm]{geometry} 
\usepackage{bm,epstopdf,epsfig,amsmath,amssymb,amsfonts,colordvi,
latexsym,comment,cancel,verbatim,url}
\usepackage{slashed}
\usepackage{cite}
\usepackage{graphicx,graphics}
\usepackage[font=md,captionskip=8pt]{subfig}
\usepackage[usenames,dvipsnames]{color}
\usepackage{setspace}

\newcommand{\nc}{\newcommand}
\renewcommand{\bar}[1]{\overline{#1}}
\renewcommand{\tilde}[1]{\widetilde{#1}}

\newcommand{\f}{\frac}
\newcommand{\be }{\begin{equation}}   \nc{\ee }{\end{equation}}
\newcommand{\bea}{\begin{eqnarray}}   \nc{\eea}{\end{eqnarray}}
\newcommand{\baa}{\begin{array}}      \nc{\eaa}{\end{array}}
\newcommand{\bit}{\begin{itemize}}    \nc{\eit}{\end{itemize}}
\newcommand{\ben}{\begin{enumerate}}  \nc{\een}{\end{enumerate}}
\nc{\bce}{\begin{center}}     \nc{\ece}{\end{center}}
\nc{\bfl}{\begin{flushright}} \nc{\efl}{\end{flushright}}
\nc{\btb}{\begin{tabular}}    \nc{\etb}{\end{tabular}}
\newcommand{\eps}{\epsilon}
\newcommand{\vp}{\varphi}
\newcommand{\tvp}{\widetilde{\varphi}}

\newcommand{\vpj }{\mbox{${\vp^\dag i\,\raisebox{2mm}{\boldmath
${}^\leftrightarrow$}\hspace{-4mm} D_\mu\,\vp}$}}
\newcommand{\vpjt}{\mbox{${\vp^\dag i\,\raisebox{2mm}{\boldmath
${}^\leftrightarrow$}\hspace{-4mm} D_\mu^{\,I}\,\vp}$}}
\def\ocal{{\cal O}}
\def\lcal{{\cal L}}
\def\p{\partial}
\def\wt{\widetilde}
\newcommand{\eq}[1]{eq.~(\ref{#1})}

\begin{document} 
\begin{flushright}
CERN-TH-2016-157\\
PSI-16-09\\
\end{flushright}

\thispagestyle{empty}

\vspace{3.5cm}

\begin{center}
{\Large {\bf Effective Field Theory with Two Higgs Doublets}}

\vspace{1cm}

{\bf Andreas Crivellin$^{\,a}$}, {\bf Margherita Ghezzi$^{\,a}$} and {\bf  Massimiliano Procura$^{\,b, \,\dagger}$}

\bigskip

{\small  $^a$ Paul Scherrer Institut, CH-5232 Villigen PSI, Switzerland}

{\small $^b$ Theoretical Physics Department, CERN, Geneva, Switzerland}
\end{center}

\vspace{10mm}

\begin{abstract}
In this article we extend the effective field theory framework describing new physics effects to the case where the underlying low-energy theory is a Two-Higgs-Doublet model. We derive a complete set of independent operators up to dimension six assuming a $Z_2$-invariant CP-conserving Higgs potential. The effects on Higgs and gauge boson masses, mixing angles in the Higgs sector as well as couplings to fermions and gauge bosons are computed. At variance with the case of a single Higgs doublet, we find that pair production of SM-like Higgses, arising through dimension-six operators, is not fixed by fermion-fermion-Higgs couplings and can therefore be sizable. 
\end{abstract}

\vspace{7cm}

-----------

\vspace{-0.1cm}

{\small $\dagger$ On leave from the University of Vienna.}


\newpage


\section{Introduction \label{sec:intro}}

Adding a second $SU(2)$ doublet scalar~\cite{Lee:1973iz} represents one of the simplest possible extensions of the Standard Model (SM) of strong and electroweak (EW) interactions. Two-Higgs-Doublet models (2HDMs) have been under extensive investigation for a long time (see for example Ref.~\cite{Gunion:1989we} for an introduction and Ref.~\cite{Branco:2011iw} for a more recent review article). There are several reasons for this interest. Firstly, these models have a rather small number of free parameters, which makes phenomenological analyses quite predictive. Additional motivation for 2HDMs is provided by axion models~\cite{Kim:1986ax}, where a global $U(1)$ Peccei-Quinn symmetry is introduced to eliminate a CP-violating term in the QCD Lagrangian~\cite{Peccei:1977hh}. Such a symmetry is only possible in scenarios with at least two Higgs doublets. Furthermore, the amount of CP violation obtained through the introduction of a second Higgs doublet can be large enough to account for the baryon asymmetry of the universe~\cite{Trodden:1998ym}.
2HDMs can also explain the anomalies observed in tauonic $B$ decays~\cite{Crivellin:2012ye,Celis:2012dk,Crivellin:2015hha}. Finally, strong motivation for studying 2HDMs is provided by the Minimal Supersymmetric Standard Model (MSSM) where supersymmetry enforces the introduction of a second Higgs doublet~\cite{Fayet:1976cr,Fayet:1976et,Haber:1984rc} due to the holomorphicity of the superpotential. 

According to present collider bounds, the additional Higgs bosons contained in these models are still allowed to have masses around the EW scale~\cite{Carena:2013ytb,PDG}\footnote{An exception is the 2HDM of type II where the bound from $b\to s\gamma$ forces the charged Higgs mass (which only differs from the other Higgs masses by terms of order $v^2$) to be larger than 400 GeV~\cite{Misiak:2015xwa}.}. Therefore, in an effective field theory (EFT) approach, it is natural to consider the additional Higgs doublet as a dynamical degree of freedom like the SM fields. In this article, we shall consider 2HDMs as EFTs valid up to a high-energy scale $\Lambda \gg M_H$ where additional dynamical degrees of freedom enter. The MSSM with heavy SUSY partners but light Higgs doublets is one example of such a theory, but also the $L_\mu-L_\tau$ model of Refs.~\cite{Crivellin:2015mga,Crivellin:2015lwa} reduces to a 2HDM if the $Z^\prime$ and the $L_\mu-L_\tau$-breaking singlet are heavy.

In general, any theory of new physics (NP) super-seeding the 2HDM at higher energies must satisfy the following requirements (in close analogy to the SM case):
\begin{itemize}
\item[({\it i})] Its gauge group contains the SM gauge group~$SU(3)_C\times SU(2)_L\times U(1)_Y$ as a subgroup.
\item[({\it ii})] It contains two Higgs doublets as dynamical degrees of freedom, either as fundamental or composite fields.
\item[({\it iii})] At low energies it reproduces the 2HDM, barring the existence of weakly coupled {\em light} particles, like axions or sterile neutrinos (in which case the EFT should include in addition these particles as dynamical degrees of freedom).
\end{itemize}

In this approach, heavier NP particles are integrated out and their effects are parameterized in terms of Wilson coefficients of higher-dimension operators suppressed by inverse powers of $\Lambda$. In our 2HDM case we have
\be 
 \label{eqn:Leff} 
  \lcal_{\mathrm 2HDM} =   \lcal_{\mathrm 2HDM}^{(4)}   + \f{1}{\Lambda }  \sum_{k} C_{k}^{(5)} Q_{k}^{(5)} + \f{1}{\Lambda^2} \sum_{k} C_k^{(6)} Q_k^{(6)}  + \ocal\!\left(\f{1}{\Lambda^3}\right)\,.
\ee
Here $\lcal_{\mathrm 2HDM}^{(4)}$ is the standard renormalizable 2HDM Lagrangian (to be specified in the next section) which contains only dimension-two and dimension-four operators. $Q_{k}^{(5)}$ generalize the Weinberg operator~\cite{Weinberg:1979sa} giving rise to neutrino masses and $Q_k^{(6)}$ denote the dimension-six operators. $C_k^{(5)}$ and $C_k^{(6)}$ are their dimensionless Wilson coefficients. In this paper we neglect the effects of operators of dimension seven and higher, which are suppressed by at least three powers of $\Lambda$.

The EFT approach to parameterize NP effects through higher-dimension operators built with SM fields (SM-EFT) has been used for a long time~\cite{Buchmuller:1985jz,Burges:1983zg,Leung:1984ni,Hagiwara:1993qt}. More recently, a complete and minimal basis of dimension-six effective operators in the SM has been established~\cite{GIMR} and the SM-EFT has received a lot of interest concerning its phenomenological applications (see for example \cite{Degrande:2012wf,Trott:2014dma,Masso:2014xra,Henning:2014wua,Contino:2016jqw}), mainly in the context of Higgs physics \cite{Giudice:2007fh,Contino:2013kra,Alloul:2013naa,Contino:2014aaa,Chen:2013kfa,Englert:2014uua,Biekoetter:2014jwa,Bonnet:2011yx,Bonnet:2012nm} but also flavor physics \cite{Crivellin:2013hpa,Crivellin:2014cta,Crivellin:2014zpa,Pruna:2014asa,Bhattacharya:2014wla,Alonso:2014csa,Buras:2014fpa,Alonso:2015sja,Calibbi:2015kma,Aebischer:2015fzz}. The possibility to extract constraints on the Wilson coefficients both from Higgs measurements at the LHC and from electroweak precision observables has been exploited~\cite{Berthier:2015oma,Ellis:2014dva,Falkowski:2014tna,Low:2009di,Pomarol:2013zra,Berthier:2015gja,Englert:2015hrx,Cirigliano:2016nyn} and an effort is currently made to perform the SM-EFT analysis at next-to-leading order in perturbation theory~\cite{Grojean:2013kd,Jenkins:2013zja,Jenkins:2013wua,Alonso:2013hga,Grober:2015cwa,Elias-Miro:2013mua,Elias-Miro:2013gya,Passarino:2012cb,Ghezzi:2015vva,Hartmann:2015oia,Hartmann:2015aia,Gauld:2015lmb}.

The purpose of this paper is to extend the SM-EFT approach to the case of two Higgs doublets. 2HDMs are considered here as low-energy theories and using our framework the effects of heavier NP particles can be clarified and studied in a systematic way. For example, effective operators in 2HDMs have been recently considered to account for a diphoton excess in LHC data~\cite{Aaboud:2016tru,Khachatryan:2016hje} with the outcome that an explanation requires to extend the field content of pure 2HDMs~\cite{Altmannshofer:2015xfo, Gupta:2015zzs,Bizot:2015qqo,Han:2016bvl,Angelescu:2015uiz,DiChiara:2015vdm,Low:2015qep,Moretti:2015pbj,Bertuzzo:2016fmv,Hernandez:2016rbi}\footnote{Prior to the LHC diphoton excess, effective operators for two-photon processes in 2HDMs were already discussed in Ref.~\cite{Perez:1995dc}.}. 

This article is organized as follows. In Sec.~\ref{sec:notation} we will introduce our notation and conventions. The complete list of operators up to dimension six before spontaneous EW symmetry breaking will be given in Sec.~\ref{operators}. We will then discuss EW symmetry breaking and the definition of the physical basis with diagonal mass matrices in Sec.~\ref{PhysicalBasis}. In Sec.~\ref{sec:conclusions} we will conclude and point out an interesting phenomenological application of our formalism.

\section{Notation and conventions \label{sec:notation}}

\begin{table}[t]
\bce \btb{|c|ccccc|c|}
\hline &&&&&&\\[-3mm]
& \multicolumn{5}{|c|}{fermions} & scalars \\[1mm]\hline&&&&&&\\[-3mm]
field & $l^j_{Lp}$ & $e_{Rp}$ & $q^{\alpha j}_{Lp}$ & $u^\alpha_{Rp}$ & $d^\alpha_{Rp}$ & $\vp_1^j, \vp_2^j$ \\[1mm]
\hline&&&&&&\\[-3mm]
hypercharge $Y$ & $-\dfrac{1}{2}$ & $-1$ & $\dfrac{1}{6}$ & $\dfrac{2}{3}$ & $-\dfrac{1}{3}$ & $\dfrac{1}{2}$ \\[2mm]
\hline 
\etb \ece
\caption{The matter content of the 2HDM. $l$ ($e$) is the lepton doublet (singlet), $u$ and $d$ the right-handed up and down quark singlets and $q$ the quark doublet. Here, $j=1,2$,~$\alpha=1,2,3$,~and~$p=1,2,3$ stand for isospin, color and generation indices, respectively.
\label{tab:matter}}
\end{table}

In this section we establish our notation and conventions following Ref.~\cite{GIMR} and Ref.~\cite{Branco:2011iw}. The renormalizable 2HDM Lagrangian before spontaneous EW symmetry breaking, reads
\begin{eqnarray} \label{2HDMlag4}
\lcal_{\mathrm 2HDM}^{(4)} &=& -\f14 G_{\mu\nu}^A G^{A\mu\nu} -\f14 W_{\mu\nu}^I W^{I\mu\nu}
-\f14 B_{\mu\nu}   B^{\mu\nu} \nonumber \\
&+& \left( D_\mu \vp_1 \right)^\dag \left( D^\mu \vp_1 \right) + \left( D_\mu \vp_2 \right)^\dag \left( D^\mu \vp_2 \right) \nonumber \\ 
   &-& V(\vp_1,\vp_2) + i  \left( \bar l {D\!\!\!\!\slash}\, l + \bar q {D\!\!\!\!\slash}\, q + \bar u {D\!\!\!\!\slash}\, u + \bar d {D\!\!\!\!\slash}\, d  \right) + {\cal L}_Y \,,
\end{eqnarray}
where $\vp_1$ and $\vp_2$ are the two Higgs doublets, $A=1 \dots 8$ ($I=1 \dots 3$) labels the $SU(3)$ ($SU(2)$) gauge bosons while $B_{\mu\nu}$ is the hypercharge field strength tensor. Note that we omitted the QCD $\theta$-term here. The fermion fields and their charges and representations are shown in Tab.~\ref{tab:matter}. For the sake of simplicity, chirality indices are suppressed in the following. The conventions for covariant derivatives are fixed {\it e.g.} by
\be \label{D-sign}
\left( D_\mu q \right)_{\alpha j} = \left( \p_\mu + i g_s T^A_{\alpha\beta} G^A_\mu 
+ i g S^I_{jk} W^I_\mu + i g' Y_q B_\mu \right) q^{\beta k}\,.
\ee
with $T^{A}=\f12 \lambda^A$ and $S^{I}=\f12 \tau^I$ denoting the $SU(3)$ and $SU(2)$ generators,
and $\lambda^A$ ($\tau^I$) the Gell-Mann (Pauli) matrices. It is useful to define the following Hermitian derivative terms:
\be \label{Dfb}
\vpj \equiv i \vp^\dag D_\mu \vp - i \left(D_\mu \vp \right)^\dag \vp
\mbox{\hspace{5mm} and \hspace{5mm}} 
\vpjt \equiv i \vp^\dag \tau^I D_\mu \vp - i \left( D_\mu \vp \right)^\dag \tau^I \vp\,.
\ee
The gauge field strength tensors are given by 
\bea 
G_{\mu\nu}^A &=& \p_\mu G_\nu^A - \p_\nu G_\mu^A - g_s f^{ABC} G_\mu^B G_\nu^C\,, \nonumber \\ 
W_{\mu\nu}^I &=& \p_\mu W_\nu^I - \p_\nu W_\mu^I - g \eps^{IJK} W_\mu^J W_\nu^K\,,\nonumber \\ 
B_{\mu\nu}   &=& \p_\mu B_\nu - \p_\nu B_\mu \,,
\eea
while
$\wt X_{\mu\nu} = \f12 \eps_{\mu\nu\rho\sigma} X^{\rho\sigma}$,
with $\eps_{0123}=+1$, denotes the dual tensor ($X = \{G^A, W^I, B\}$).

We consider a CP-conserving scalar potential~\cite{Branco:2011iw}
\bea \label{potential}
V(\vp_1,\vp_2) &=&
m^2_{11}\, \vp_1^\dagger \vp_1
+ m^2_{22}\, \vp_2^\dagger \vp_2 -
 m^2_{12}\, \left(\vp_1^\dagger \vp_2 + \vp_2^\dagger \vp_1\right)
+ \frac{\lambda_1}{2} \left( \vp_1^\dagger \vp_1 \right)^2
+ \frac{\lambda_2}{2} \left( \vp_2^\dagger \vp_2 \right)^2
\nonumber  \\ & & 
+ \lambda_3\, \vp_1^\dagger \vp_1\, \vp_2^\dagger \vp_2
+ \lambda_4\, \vp_1^\dagger \vp_2\, \vp_2^\dagger \vp_1
+ \frac{\lambda_5}{2} \left[
\left( \vp_1^\dagger\vp_2 \right)^2
+ \left( \vp_2^\dagger\vp_1 \right)^2 \right] \,
\eea
with a $Z_2$ symmetry~\footnote{Note that assigning Peccei-Quinn charges to Higgs doublets and fermions has the same effect on the potential and the Yukawa couplings as imposing a $Z_2$ symmetry.} (softly broken by a dimension-two term) preventing the existence of terms with odd powers of $\vp_1$ and $\vp_2$. Here all parameters are assumed to be real~\footnote{For an analysis of the conditions for a CP-conserving Higgs sector we refer to~\cite{Gunion:2002zf}.}. In order to study fluctuations around the VEVs that minimize the potential~\footnote{Since no CP violation is involved, both $v_1$ and $v_2$ can be taken to be real.}, the two complex scalar fields are parameterized as
\be
\vp_a = \left( \begin{array}{c} \phi_a^+ \\
\left(v_a +  \rho_a + i \eta_a \right) \left/ \sqrt{2} \right.
\end{array} \right),
\quad a= 1, 2\,.
\ee
The Lagrangian for the mass terms of the CP-odd ($\eta_a$), CP-even ($\rho_a$) and charged ($\phi_a^+$) Higgses is
\begin{eqnarray}
\vphantom{00}\hspace{-.7cm}
L_{{M_H}}^{(4)} \!\!&=&\!\! \frac{1}{2}{\mkern 1mu} {\left( {\begin{array}{*{20}{c}}
{{\eta _1}}\\
{{\eta _2}}
\end{array}} \right)^T}\!m_\eta^2\!\left( {\begin{array}{*{20}{c}}
{{\eta _1}}\\
{{\eta _2}}
\end{array}} \right) + {\left( {\begin{array}{*{20}{c}}
{\phi _1^ - }\\
{\phi _2^ - }
\end{array}} \right)^T}\!m_{{\phi ^ \pm }}^2\!\left( {\begin{array}{*{20}{c}}
{\phi _1^ + }\\
{\phi _2^ + }
\end{array}} \right) + {\mkern 1mu} \frac{1}{2}{\left( {\begin{array}{*{20}{c}}
{{\rho _1}}\\
{{\rho _2}}
\end{array}} \right)^T}\!m_\rho ^2\!\left( {\begin{array}{*{20}{c}}
{{\rho _1}}\\
{{\rho _2}}
\end{array}} \right)
\end{eqnarray}
with
\begin{eqnarray}
m_\eta ^2 \!\!&=&\!\! \left( {{v_1}{v_2}{\lambda _5} - m_{12}^2} \right)\left( {\begin{array}{*{20}{c}}
{ - \dfrac{{{v_2}}}{{{v_1}}}}&1\\
1&{ - \dfrac{{{v_1}}}{{{v_2}}}}
\end{array}} \right)\\
m_\rho ^2 \!\!&=&\!\! \left( {\begin{array}{*{20}{c}}
{{\lambda _1}{\mkern 1mu} v_1^2 + m_{12}^2{\mkern 1mu} \dfrac{{{v_2}}}{{{v_1}}}}&{{v_1}{v_2}\left( {{\lambda _3} + {\lambda _4} + {\lambda _5}} \right) - m_{12}^2}\\
{{v_1}{v_2}\left( {{\lambda _3} + {\lambda _4} + {\lambda _5}} \right) - m_{12}^2}&{{\lambda _2}{\mkern 1mu} v_2^2 + m_{12}^2{\mkern 1mu} \dfrac{{{v_1}}}{{{v_2}}}}
\end{array}} \right)\\
m_{{\phi ^ \pm }}^2 \!\!&=&\!\! \left[ {\frac{{{v_1}{v_2}}}{2}\left( {{\lambda _4} + {\lambda _5}} \right) - m_{12}^2} \right]\left( {\begin{array}{*{20}{c}}
{ - \dfrac{{{v_2}}}{{{v_1}}}}&1\\
1&{ - \dfrac{{{v_1}}}{{{v_2}}}}
\end{array}} \right)
\end{eqnarray}
where we eliminated $m_{11}^2$ and $m_{22}^2$ by the minimization conditions. The charged and CP-odd mass matrices have both one vanishing eigenvalue, which corresponds to the Goldstone bosons giving masses to the $W$ and the $Z$, and a non-zero eigenvalue named $m_{H^\pm}^2$ and $m_A^2$, respectively. Both matrices are diagonalized by the same angle $\beta$ defined as
\begin{equation}
	\tan \beta \equiv v_2/v_1\,.
\end{equation}
Another independent rotation angle, called $\alpha$, enters the definition of the CP-even mass eigenstates $h$ and $H$, with eigenvalues $m_h$ and $m_H$ respectively:
\bea
h &=& \rho_1 \sin{\alpha} - \rho_2 \cos{\alpha}\,, \nonumber
\\ 
H &=& - \rho_1 \cos{\alpha} - \rho_2 \sin{\alpha}\,
\label{eq:alpha}
\eea
where usually the lighter one is identified with the SM Higgs, with mass $m_h \approx 125\,{\rm GeV}$.

Let us finally turn to the Yukawa part of the Lagrangian in \eq{2HDMlag4}:
\begin{equation}
	{\cal L}_Y = -Y_1^e\,\bar l{\varphi _1}e - Y_2^e\,\bar l{\varphi _2}e - Y_1^d\,\bar q{\varphi _1}d - Y_2^d\,\bar q{\varphi _2}d - Y_1^u\,\bar q{{\tilde \varphi }_1}u - Y_2^u\,\bar q{{\tilde \varphi }_2}u + h.c.
\end{equation}
Here we suppressed fermion flavor indices and defined $\tvp^j = \eps_{jk}(\vp^k)^\star$, using the totally antisymmetric $\eps_{jk}$ with $\eps_{12}=+1$. The Yukawa couplings $Y^f_{1,2}$ are understood to be $3 \times 3$ matrices in flavor space. If $Y_1^f$ and $Y_2^f$ are simultaneously non-zero, in general flavor changing neutral currents arise~\cite{Bjorken:1977vt,McWilliams:1980kj,Cheng:1987rs}. However, there are four 2HDMs with natural flavor conservation (see Table~\ref{2HDMs}) where only one of these couplings is present. As for CP-conservation in the potential, this can be achieved by an appropriate $Z_2$ charge assignment to right-handed fermions. When we discuss the extension to dimension six, we will assume that the terms are made $Z_2$-invariant in the same way.

\begin{table}
\renewcommand{\arraystretch}{1.3}
\centering
\begin{tabular}{|c|c|c|c|}
\hline
{{$\rm{model}$}}& {{$u_R$}}&{{$d_R$}}&{{$e_R$}} \\
\hline
{$\rm{Type\; I}$}& {{$\varphi _2$}}&{{$\varphi _2$}}&{{$\varphi _2$}} \\
{{$\rm{Type\; II}$}}& {{$\varphi _2$}}&{{$\varphi _1$}}&{{$\varphi _1$}} \\
{{$\rm{Lepton - specific}$}}& {{$\varphi _2$}}&{{$\varphi _2$}}&{{$\varphi _1$}} \\
{{$\rm{Flipped}$}}& {{$\varphi _2$}}&{{$\varphi _1$}}&{{$\varphi _2$}}\\
\hline
\end{tabular}
\caption{Couplings of right-handed fermion singlets to Higgs doublets in 2HDMs with natural flavor conservation. These couplings can be enforced by an appropriate assignment of $Z_2$ (or Peccei-Quinn) charges to the scalar doublets and right-handed fermions. }
\label{2HDMs}
\end{table}
 
 
\section{Gauge invariant operators}
\label{operators}

In this section we list the independent gauge invariant operators up to dimension six in the 2HDM-EFT. They are defined before the EW symmetry breaking takes place, meaning that they are given in the interaction basis, as the mass basis is not yet defined. After EW symmetry breaking, the fermions acquire masses and also the Higgs mass matrices receive additional contributions compared to the 2HDM with dimension-four operators only. 

\begin{table}
\renewcommand{\arraystretch}{1.6}
\centering
\begin{tabular}{|c|}
\hline
$\varphi ^6$ \\
\hline
 $ Q_\varphi ^{111} = {( {\varphi _1^\dag {\varphi _1}} )^3} \hfill $ \\
 $ Q_\varphi ^{112} = {( {\varphi _1^\dag {\varphi _1}} )^2}( {\varphi _2^\dag {\varphi _2}} ) \hfill $ \\
 $ Q_\varphi ^{122} = ( {\varphi _1^\dag {\varphi _1}} ){( {\varphi _2^\dag {\varphi _2}} )^2} \hfill$\\
 $ Q_\varphi ^{222} = {( {\varphi _2^\dag {\varphi _2}} )^3} \hfill $\\ 
 $Q_\varphi ^{( {1221} )1} = ( {\varphi _1^\dag {\varphi _2}} )( {\varphi _2^\dag {\varphi _1}} )( {\varphi _1^\dag {\varphi _1}} ) \hfill$\\
 $Q_\varphi ^{( {1221} )2} = ( {\varphi _1^\dag {\varphi _2}} )( {\varphi _2^\dag {\varphi _1}} )( {\varphi _2^\dag {\varphi _2}} ) \hfill $\\
$Q_\varphi ^{( {1212} )1} = {( {\varphi _1^\dag {\varphi _2}} )^2}( {\varphi _1^\dag {\varphi _1}} ) + h.c. \hfill$\\
$Q_\varphi ^{( {1212} )2} = {( {\varphi _1^\dag {\varphi _2}} )^2}( {\varphi _2^\dag {\varphi _2}} ) + h.c. \hfill$\\ [1mm]
\hline
\end{tabular}
\caption{\small Operators in the 2HDM-EFT containing six Higgs doublets.}
\label{6higgs}
\end{table}

At dimension five the generalization of the Weinberg operator reads
\begin{eqnarray}
Q_{\nu\nu}^{11} &= (\tilde {\varphi}_1^\dagger l_p)^T C (\tilde {\varphi}_1^\dagger l_r)\,, \,\,\,\,\,
Q_{\nu\nu}^{22} &= (\tilde {\varphi}_2^\dagger l_p)^T C (\tilde {\varphi}_2^\dagger l_r)
\end{eqnarray}
where $C$ denotes the charge conjugation matrix.

The procedure we follow to obtain a complete set of independent operators at dimension six is the same as the one applied and thoroughly described in Ref.~\cite{GIMR} for SM-EFT. Obviously, the operators involving no Higgs doublets do not change compared to SM-EFT, and for them we refer the reader to Ref.~\cite{GIMR}. Using classical equations of motion, neglecting total derivatives, and imposing the constraint of vanishing total hypercharge, we derived a set of independent operators, which we classify like in the case of the SM-EFT as follows: 

\begin{itemize}
	\item $\varphi ^6$: Operators with Higgs doublets only (Table~\ref{6higgs}), which modify the Higgs potential. We assumed that these operators respect the $Z_2$ symmetry present at dimension four. 
	\item ${\varphi ^4}{D^2}$: Operators with four Higgs doublets and two derivatives (Table~\ref{4higgs2D}), which modify the kinetic terms of the Higgs fields, the Higgs-gauge boson interactions and the $W$ and $Z$ masses.
	\item ${\Psi ^2}{\varphi}X$: Operators with two fermion fields, one field strength tensor and one Higgs doublet (Table~\ref{2fermion1higgs1X}), which give rise to dipole interactions after EW symmetry breaking.
	\item ${\varphi ^2}{X^2}$: Operators with two Higgs doublets and two field strength tensors (Table~\ref{2higgs2bosons}).
	\item ${\Psi ^2}{\varphi ^2} D$: Operators with two fermions fields, two Higgs doublets and one covariant derivative (Table~\ref{2fermion2higgs1D}), which contribute to the fermion-$Z$ and fermion-$W$ couplings after EW symmetry breaking.
	\item ${\Psi ^2}{\varphi ^3}$: Operators containing two fermion fields and three Higgs doublets (Table~\ref{2fermion3higgs}), which modify the relation between fermion masses and Higgs-fermion couplings.
\end{itemize}

\begin{table}
\renewcommand{\arraystretch}{2.0}
\hspace{-1.0cm}
\begin{small}
\begin{tabular}{|c|c|}
\hline
\multicolumn{2}{|c|} {${\varphi ^4}{D^2}$} \\
\hline
$ \square $ & $ \varphi D $ \\
\hline
$ Q_\square^{1\left( 1 \right)} = ( {\varphi _1^\dag {\varphi _1}} )\square( {\varphi _1^\dag {\varphi _1}} ) \hfill  $
& ${Q_{\varphi D}^{\left( 1 \right)11\left( 1 \right)} = \left[ {{{\left( {{D_\mu }\varphi _1^{}} \right)}^\dag }{\varphi _1}} \right]\left[ {\varphi _1^\dag \left( {{D^\mu }{\varphi _1}} \right)} \right]}\;\;\;\; {Q_{\varphi D}^{\left( 1 \right)21\left( 2 \right)} = \left[ {{{\left( {{D_\mu }\varphi _1^{}} \right)}^\dag }{\varphi _2}} \right]\left[ {\varphi _1^\dag \left( {{D^\mu }{\varphi _2}} \right)} \right] + h.c.}$ \\ 
$ Q_\square^{2( 2 )} = ( {\varphi _2^\dag {\varphi _2}} )\square( {\varphi _2^\dag {\varphi _2}} ) \hfill $  
& ${Q_{\varphi D}^{\left( 2 \right)22\left( 2 \right)} = \left[ {{{\left( {{D_\mu }\varphi _2^{}} \right)}^\dag }{\varphi _2}} \right]\left[ {\varphi _2^\dag \left( {{D^\mu }{\varphi _2}} \right)} \right]}\;\;\;\;{Q_{\varphi D}^{\left( 1 \right)12\left( 2 \right)} = \left[ {{{\left( {{D_\mu }\varphi _1^{}} \right)}^\dag }{\varphi _1}} \right]\left[ {\varphi _2^\dag \left( {{D^\mu }{\varphi _2}} \right)} \right] + h.c.}$\\
$ Q_\square^{1 2 } = ( {\varphi _1^\dag {\varphi _1}} )\square( {\varphi _2^\dag {\varphi _2}} ) \hfill $ 
& ${Q_{\varphi D}^{\left( 1 \right)22\left( 1 \right)} = \left[ {{{\left( {{D_\mu }\varphi _1^{}} \right)}^\dag }{\varphi _2}} \right]\left[ {\varphi _2^\dag \left( {{D^\mu }{\varphi _1}} \right)} \right]}\;\;\;\;{Q_{\varphi D}^{12\left( {12} \right)} = \left[ {\varphi {{_1^{}}^\dag }{\varphi _2}} \right]\left[ {{{\left( {{D_\mu }\varphi _1^{}} \right)}^\dag }\left( {{D^\mu }{\varphi _2}} \right)} \right] + h.c.}$\\ 
$ \;\;\;\;\;\;\;\;+ ( {\varphi _2^\dag {\varphi _2}} )\square( {\varphi _1^\dag {\varphi _1}} ) \hfill $
& ${Q_{\varphi D}^{\left( 2 \right)11\left( 2 \right)} = \left[ {{{\left( {{D_\mu }\varphi _2^{}} \right)}^\dag }{\varphi _1}} \right]\left[ {\varphi _1^\dag \left( {{D^\mu }{\varphi _2}} \right)} \right]}\;\;\;\;{Q_{\varphi D}^{12\left( {21} \right)} = \left[ {\varphi {{_1^{}}^\dag }{\varphi _2}} \right]\left[ {{{\left( {{D_\mu }\varphi _2^{}} \right)}^\dag }\left( {{D^\mu }{\varphi _1}} \right)} \right] + h.c.}$\hfill\\ [1mm]
\hline
\end{tabular}
\end{small}
\caption{ \small Operators with four Higgs doublets and two derivatives. }\label{4higgs2D}
\end{table}

As in the case of the dimension-four Lagrangian, we assume that the discrete $Z_2$ symmetry for operators involving Higgs and fermion fields is restored by an appropriate charge assignment to right-handed fermions. Concerning Table~\ref{4higgs2D}, we stress that, as in the case of one Higgs doublet, operators with derivatives acting on two conjugated or two unconjugated fields are not independent.

\begin{table}
\renewcommand{\arraystretch}{1.8}
\centering
\begin{tabular}{|c|c|c|}
\hline
\multicolumn{3}{|c|} {${\Psi ^2}{\varphi}X$} \\
\hline
$G$ & $W$ & $B$ \\ 
\hline
$Q_{dG}^1 = ({{\bar q}_p}{\sigma ^{\mu \nu }}{T^A}{d_r}){\varphi _1}{\kern 1pt} G_{\mu \nu }^A \hfill $ & $Q_{dW}^1 = ({{\bar q}_p}{\sigma ^{\mu \nu }}{d_r}){\tau ^I}{\varphi _1}{\kern 1pt} W_{\mu \nu }^I \hfill $ & $ Q_{dB}^1 = ({{\bar q}_p}{\sigma ^{\mu \nu }}{d_r})\varphi {{\kern 1pt} _1}{B_{\mu \nu }} \hfill $ \\
$ Q_{dG}^2 = ({{\bar q}_p}{\sigma ^{\mu \nu }}{T^A}{d_r})\varphi {{\kern 1pt} _2}G_{\mu \nu }^A \hfill $ & $ Q_{dW}^2 = ({{\bar q}_p}{\sigma ^{\mu \nu }}{d_r}){\tau ^I}{\varphi _2}{\kern 1pt} W_{\mu \nu }^I \hfill $ & $  Q_{dB}^2 = ({{\bar q}_p}{\sigma ^{\mu \nu }}{d_r})\varphi {{\kern 1pt} _2}{B_{\mu \nu }}\; \hfill $ \\
$ Q_{uG}^1 = ({{\bar q}_p}{\sigma ^{\mu \nu }}{T^A}{u_r})\tilde \varphi {{\kern 1pt} _1}{G_{\mu \nu }^A} \hfill $ & $ Q_{uW}^1 = ({{\bar q}_p}{\sigma ^{\mu \nu }}{u_r}){\tau ^I}{{\tilde \varphi }_1}{\kern 1pt} W_{\mu \nu }^I \hfill $ & $ Q_{uB}^1 = ({{\bar q}_p}{\sigma ^{\mu \nu }}{u_r})\tilde \varphi {{\kern 1pt} _1}{B_{\mu \nu }} \hfill $ \\
$ Q_{uG}^2 = ({{\bar q}_p}{\sigma ^{\mu \nu }}{T^A}{u_r})\tilde \varphi {{\kern 1pt} _2}{G_{\mu \nu }^A} \hfill $ & $ Q_{uW}^2 = ({{\bar q}_p}{\sigma ^{\mu \nu }}{u_r}){\tau ^I}{{\tilde \varphi }_2}{\kern 1pt} W_{\mu \nu }^I \hfill $ & $ Q_{uB}^2 = ({{\bar q}_p}{\sigma ^{\mu \nu }}{u_r})\tilde \varphi {{\kern 1pt} _2}{B_{\mu \nu }} \hfill $ \\
 & $ Q_{eW}^1 = ({{\bar l}_p}{\sigma ^{\mu \nu }}{e_r}){\tau ^I}{\varphi _1}{W_{\mu \nu }^I} \hfill  $ & $ Q_{eB}^1 = ({{\bar l}_p}{\sigma ^{\mu \nu }}{e_r}){\varphi _1}{B_{\mu \nu }} \hfill $ \\
  & $ Q_{eW}^2 = ({{\bar l}_p}{\sigma ^{\mu \nu }}{e_r}){\tau ^I}{\varphi _2}{W_{\mu \nu }^I} \hfill $ & $ Q_{eB}^2 = ({{\bar l}_p}{\sigma ^{\mu \nu }}{e_r}){\varphi _2}{B_{\mu \nu }}\; \hfill $ \\ [1mm]
\hline
\end{tabular}
\caption{ \small Operators containing two fermion fields, one Higgs doublet and a field strength tensor. Here $\sigma^{\mu \nu}= i\, [\gamma^\mu, \gamma^\nu]/2.$ }
\label{2fermion1higgs1X}
\end{table}

\begin{table}
\renewcommand{\arraystretch}{1.8}
\centering
\begin{tabular}{|c|c|}
\hline
\multicolumn{2}{|c|} {${\varphi ^2}{X^2}$ } \\
\hline
 $GG, WW, BB$ & $WB$ \\
 \hline
 $ Q_{\varphi X}^{11} = ( {\varphi _1^\dag {\varphi _1}{\mkern 1mu} } )X_{\mu \nu }{X^{\mu \nu }} \hfill $  & $ Q_{\varphi WB}^{11} = ( {\varphi _1^\dag {\tau ^I}{\varphi _1}{\mkern 1mu} } )W_{\mu \nu }^I{B^{\mu \nu }} \hfill $\\
 $Q_{\varphi X}^{22} = ( {\varphi _2^\dag {\varphi _2}{\mkern 1mu} } )X_{\mu \nu }{X^{\mu \nu }} \hfill $  & $Q_{\varphi WB}^{22} = ( {\varphi _2^\dag {\tau ^I}{\varphi _2}{\mkern 1mu} } )W_{\mu \nu }^I{B^{\mu \nu }} \hfill $\\
 $ Q_{\varphi \tilde X}^{11} = ( {\varphi _1^\dag {\varphi _1}{\mkern 1mu} } )\tilde X_{\mu \nu }{X^{\mu \nu }} \hfill$  & $Q_{\varphi \tilde W B}^{11} = ( {\varphi _1^\dag  {\tau ^I}{\varphi _1}{\mkern 1mu} } )\tilde W_{\mu \nu }^I{B^{\mu \nu }} \hfill $\\
 $ Q_{\varphi \tilde X}^{22} = ( {\varphi _2^\dag {\varphi _2}{\mkern 1mu} } )\tilde X_{\mu \nu }{X^{\mu \nu }} \hfill $ & $Q_{\varphi \tilde W B}^{22} = ( {\varphi _2^\dag {\tau ^I} {\varphi _2}{\mkern 1mu} } )\tilde W_{\mu \nu }^I{B^{\mu \nu }} \hfill $\\ [1mm]
\hline
\end{tabular}
\caption{\small Operators with two scalar fields and two field strength tensors. $X$ denotes $G^A$, $W^I$ or $B$.}\label{2higgs2bosons}
\end{table}

\begin{table}
\renewcommand{\arraystretch}{1.4}
\centering
\begin{tabular}{|c|c|}
\hline
\multicolumn{2}{|c|} {${\Psi ^2}{\varphi ^2} D$} \\
\hline
 ${\left( 1 \right)}$&${\left( 3 \right)} $ \\
 \hline
 $Q_{\varphi ud}^1 = i( {\tilde\varphi _1^\dag i{{\overset{\lower0.5em\hbox{$\smash{\scriptscriptstyle\leftrightarrow}$}} {D} }_\mu }{\varphi _1}} )( {{{\bar u}_p}{\gamma ^\mu }{d_r}} ) \hfill $ & \\
  $Q_{\varphi ud}^2 = i( {\tilde\varphi _2^\dag i{{\overset{\lower0.5em\hbox{$\smash{\scriptscriptstyle\leftrightarrow}$}} {D} }_\mu }{\varphi _2}} )( {{{\bar u}_p}{\gamma ^\mu }{d_r}} ) \hfill$ & \\ 

$  Q_{\varphi l}^{(1)1} = ( {\varphi _1^\dag i{{\overset{\lower0.5em\hbox{$\smash{\scriptscriptstyle\leftrightarrow}$}} {D} }_\mu }{\varphi _1}} )( {{{\bar l}_p}{\gamma ^\mu }{l_r}} ) \hfill $ & $  Q_{\varphi l}^{(3)1} = ( {\varphi _1^\dag i \,\raisebox{2mm}{\boldmath
${}^\leftrightarrow$}\hspace{-4mm} D_\mu ^I{\varphi _1}} )( {{{\bar l}_p}{\tau ^I}{\gamma ^\mu }{l_r}} ) \hfill $\\

$  Q_{\varphi l}^{(1)2} = ( {\varphi _2^\dag i{{\overset{\lower0.5em\hbox{$\smash{\scriptscriptstyle\leftrightarrow}$}} {D} }_\mu } {\varphi _2}} )( {{{\bar l}_p}{\gamma ^\mu }{l_r}} ) \hfill $ &  $ Q_{\varphi l}^{(3)2} = ( {\varphi _2^\dag i\,\raisebox{2mm}{\boldmath
${}^\leftrightarrow$}\hspace{-4mm} D_\mu  ^I{\varphi _2}} )( {{{\bar l}_p}{\tau ^I}{\gamma ^\mu }{l_r}} ) \hfill $\\ 

$  Q_{\varphi e}^1 = ( {\varphi _1^\dag i{{\overset{\lower0.5em\hbox{$\smash{\scriptscriptstyle\leftrightarrow}$}} {D} }_\mu }{\varphi _1}} )( {{{\bar e}_p}{\gamma ^\mu }{e_r}} ) \hfill $ & \\
 $ Q_{\varphi e}^2 = ( {\varphi _2^\dag i{{\overset{\lower0.5em\hbox{$\smash{\scriptscriptstyle\leftrightarrow}$}} {D} }_\mu }{\varphi _2}} )( {{{\bar e}_p}{\gamma ^\mu }{e_r}} ) \hfill $ & \\ 

 $ Q_{\varphi q}^{(1)1} = ( {\varphi _1^\dag i{{\overset{\lower0.5em\hbox{$\smash{\scriptscriptstyle\leftrightarrow}$}} {D} }_\mu }{\varphi _1}} )( {{{\bar q}_p}{\gamma ^\mu }{q_r}} ) \hfill $ & $ Q_{\varphi q}^{(3)1} = ( {\varphi _1^\dag i \,\raisebox{2mm}{\boldmath
${}^\leftrightarrow$}\hspace{-4mm} D_\mu ^I{\varphi _1}} )( {{{\bar q}_p}{\tau ^I}{\gamma ^\mu }{q_r}} ) \hfill $ \\

 $ Q_{\varphi q}^{(1)2} = ( {\varphi _2^\dag i{{\overset{\lower0.5em\hbox{$\smash{\scriptscriptstyle\leftrightarrow}$}} {D} }_\mu }{\varphi _2}} )( {{{\bar q}_p}{\gamma ^\mu }{q_r}} ) \hfill $ & $ Q_{\varphi q}^{(3)2} = ( {\varphi _2^\dag i \,\raisebox{2mm}{\boldmath
${}^\leftrightarrow$}\hspace{-4mm} D_\mu ^I{\varphi _2}} )( {{{\bar q}_p}{\tau ^I}{\gamma ^\mu }{q_r}} ) \hfill $ \\ 

 $ Q_{\varphi u}^1 = ( {\varphi _1^\dag i{{\overset{\lower0.5em\hbox{$\smash{\scriptscriptstyle\leftrightarrow}$}} {D} }_\mu }{\varphi _1}} )( {{{\bar u}_p}{\gamma ^\mu }{u_r}} ) \hfill $ & \\
 $ Q_{\varphi u}^2 = ( {\varphi _2^\dag i{{\overset{\lower0.5em\hbox{$\smash{\scriptscriptstyle\leftrightarrow}$}} {D} }_\mu }{\varphi _2}} )( {{{\bar u}_p}{\gamma ^\mu }{u_r}} ) \hfill $ & \\ 

 $ Q_{\varphi d}^1 = ( {\varphi _1^\dag i{{\overset{\lower0.5em\hbox{$\smash{\scriptscriptstyle\leftrightarrow}$}} {D} }_\mu }{\varphi _1}} )( {{{\bar d}_p}{\gamma ^\mu }{d_r}} ) \hfill $ & \\
 $ Q_{\varphi d}^2 = ( {\varphi _2^\dag i{{\overset{\lower0.5em\hbox{$\smash{\scriptscriptstyle\leftrightarrow}$}} {D} }_\mu }{\varphi _2}} )( {{{\bar d}_p}{\gamma ^\mu }{d_r}} ) \hfill $ & \\ [1mm]
\hline
\end{tabular}
\caption{\small Operators in the 2HDM-EFT containing two fermions, two Higgs doublets and a covariant derivative. The superscripts (3), (1) label fermion bilinears transforming as an $SU(2)$ triplet or singlet, respectively.} 
\label{2fermion2higgs1D}
\end{table}

\begin{table}
\renewcommand{\arraystretch}{1.4}
\centering
\begin{tabular}{|c|c|c|}
\hline
\multicolumn{3}{|c|} {${\Psi ^2}{\varphi ^3}$} \\
\hline
$e$ & $d$ & $u$ \\
\hline
$ Q_{e\varphi }^{111} = ( {{{\bar l}_p}{e_r}{\varphi _1}} )( {\varphi _1^\dag {\varphi _1}} ) \hfill $ & $ Q_{d\varphi }^{111} =( {{{\bar q}_p}{d_r}{\varphi _1}} )( {\varphi _1^\dag {\varphi _1}} ) \hfill $ & $ Q_{u\varphi }^{111} = ( {{{\bar q}_p}{u_r}{{\tilde \varphi }_1}} )( {\varphi _1^\dag {\varphi _1}} ) \hfill $ \\
$ Q_{e\varphi }^{122} = ( {{{\bar l}_p}{e_r}{\varphi _1}} )( {\varphi _2^\dag {\varphi _2}} ) \hfill $ & $ Q_{d\varphi }^{122} = ( {{{\bar q}_p}{d_r}{\varphi _1}} )( {\varphi _2^\dag {\varphi _2}} ) \hfill $ & $ Q_{u\varphi }^{122} = ( {{{\bar q}_p}{u_r}{{\tilde \varphi }_1}} )( {\varphi _2^\dag {\varphi _2}} ) \hfill $ \\
$ Q_{e\varphi }^{222} = ( {{{\bar l}_p}{e_r}{\varphi _2}} )( {\varphi _2^\dag {\varphi _2}} ) \hfill $ & $ Q_{d\varphi }^{222} = ( {{{\bar q}_p}{d_r}{\varphi _2}} )( {\varphi _2^\dag {\varphi _2}} ) \hfill $ & $ Q_{u\varphi }^{222} = ( {{{\bar q}_p}{u_r}{{\tilde \varphi }_2}} )( {\varphi _2^\dag {\varphi _2}} ) \hfill $ \\
$ Q_{e\varphi }^{211} = ( {{{\bar l}_p}{e_r}{\varphi _2}} )( {\varphi _1^\dag {\varphi _1}} ) \hfill $ & $ Q_{d\varphi }^{211} = ( {{{\bar q}_p}{d_r}{\varphi _2}} )( {\varphi _1^\dag {\varphi _1}} ) \hfill $ & $ Q_{u\varphi }^{211} = ( {{{\bar q}_p}{u_r}{{\tilde \varphi }_2}} )( {\varphi _1^\dag {\varphi _1}} )\; \hfill $ \\ [1mm]
\hline
\end{tabular}
\caption{\small Operators in the 2HDM-EFT containing two fermion fields and three Higgs doublets. }
\label{2fermion3higgs}
\end{table}

\section{Physical Basis}
\label{PhysicalBasis}

In this section we discuss the modifications of the Higgs potential and the relation
between the Yukawa couplings and the fermion masses induced by the dimension-six
contributions. The operators affecting Higgs kinetic terms and Higgs potential (and thus the Higgs mass
matrices) are given in Table~\ref{4higgs2D} and Table~\ref{6higgs}, respectively. The modified relations between fermion masses and Higgs-fermion couplings stem from the operators in Table~\ref{2fermion3higgs}. The kinetic terms of the gauge boson fields receive contributions from the operators in Table~\ref{2higgs2bosons}.

\subsection{Kinetic terms, Higgs and gauge boson masses}

The effect of the operators in Table~\ref{4higgs2D} on the kinetic terms of the Higgs fields amounts to
\begin{equation}
\begin{array}{l}
\arraycolsep=1.5pt\def\arraystretch{1.9}
L_{{H_{{\rm{kin}}}}}^{\left( 4 \right)+\left( 6 \right)} = \dfrac{1}{2}{\left( {\begin{array}{*{20}{c}}
\partial _\mu \rho _1\\
\partial _\mu \rho _2
\end{array}} \right)^{\!T}}\left( {\begin{array}{*{20}{c}}
{1 + \dfrac{{2\Delta _\square^{11}}}{{{\Lambda ^2}}} + \dfrac{{\Delta _{\varphi D}^{11}}}{{2{\Lambda ^2}}}}&{\dfrac{{\Delta _\square^{12}}}{{{\Lambda ^2}}} + \dfrac{{\Delta _{\varphi D}^{12}}}{{2{\Lambda ^2}}}}\\ 
{\dfrac{{\Delta _\square^{12}}}{{{\Lambda ^2}}} + \dfrac{{\Delta _{\varphi D}^{12}}}{{2{\Lambda ^2}}}}&{1 + \dfrac{{2\Delta _\square^{22}}}{{{\Lambda ^2}}} + \dfrac{{\Delta _{\varphi D}^{22}}}{{2{\Lambda ^2}}}}
\end{array}} \right)\left( {\begin{array}{*{20}{c}}
{{\partial _\mu }{\rho _1}}\\
{{\partial _\mu }{\rho _2}}
\end{array}} \right)\\ \\
\arraycolsep=1.5pt\def\arraystretch{1.9}
 \qquad \quad +\, \dfrac{1}{2}{\left( {\begin{array}{*{20}{c}}
{{\partial _\mu }{\eta _1}}\\
{{\partial _\mu }{\eta _2}}
\end{array}} \right)^T}\left( {\begin{array}{*{20}{c}}
{1 + \dfrac{{\Delta _{\varphi D}^{11}}}{{2{\Lambda ^2}}}}&{\dfrac{{\Delta _{\varphi D}^{12}}}{{2{\Lambda ^2}}}}\\
{\dfrac{{\Delta _{\varphi D}^{12}}}{{2{\Lambda ^2}}}}&{1 + \dfrac{{\Delta _{\varphi D}^{22}}}{{2{\Lambda ^2}}}}
\end{array}} \right)\left( {\begin{array}{*{20}{c}}
{{\partial _\mu }{\eta _1}}\\
{{\partial _\mu }{\eta _2}}
\end{array}} \right)
\\ \\
\arraycolsep=1.5pt\def\arraystretch{1.9}
 \qquad \quad +\, {\left( {\begin{array}{*{20}{c}}
{{\partial _\mu }\phi _1^ + }\\
{{\partial _\mu }\phi _2^ + }
\end{array}} \right)^\dag }\left( {\begin{array}{*{20}{c}}
1&{\dfrac{{\Delta _{\varphi D}^ + }}{{2{\Lambda ^2}}}}\\
{\dfrac{{\Delta _{\varphi D}^ + }}{{2{\Lambda ^2}}}}&1
\end{array}} \right)\left( {\begin{array}{*{20}{c}}
{{\partial _\mu }\phi _1^ + }\\
{{\partial _\mu }\phi _2^ + }
\end{array}} \right)
\end{array}
\end{equation}
with
\begin{equation}
\renewcommand{\arraystretch}{1.6}
\begin{array}{l}
\Delta _\square^{11} =  - C_\square^{1\left( 1 \right)}v_1^2\\
\Delta _\square^{12} =  - C_\square^{1 2} {v_1}{v_2}\\
\Delta _\square^{22} =  - C_\square^{2\left( 2 \right)}v_2^2\\
\Delta _{\varphi D}^{11} =  {C_{\varphi D}^{\left( 1 \right)11\left( 1 \right)}v_1^2 + C_{\varphi D}^{\left( 1 \right)22\left( 1 \right)}v_2^2} \\
\Delta _{\varphi D}^{22} =  {C_{\varphi D}^{\left( 2 \right)11\left( 2 \right)}v_1^2 + C_{\varphi D}^{\left( 2 \right)22\left( 2 \right)}v_2^2} \\
\Delta _{\varphi D}^{12} = {v_1}{v_2}\left( {C_{\varphi D}^{\left( 1 \right)12\left( 2 \right)} + C_{\varphi D}^{\left( 1 \right)21\left( 2 \right)} + C_{\varphi D}^{12\left( {21} \right)} + C_{\varphi D}^{12\left( {12} \right)}} \right)\\
\Delta _{\varphi D}^ +  = {v_1}{v_2}\left( {C_{\varphi D}^{12\left( {21} \right)} + C_{\varphi D}^{12\left( {12} \right)}} \right)\, .
\end{array}
\end{equation}
The kinetic terms are made canonical by the shifts 
\begin{equation}
\renewcommand{\arraystretch}{1.8}
\begin{array}{l}
{\rho _1} \to {\rho _1}\left( {1 - \dfrac{{\Delta _{\varphi D}^{11} + 4\Delta _\square^{11}}}{{4{\Lambda ^2}}}} \right) - \left( {\dfrac{{\Delta _{\varphi D}^{12} + 4\Delta _\square^{12}}}{{4{\Lambda ^2}}}} \right){\rho _2}\\
{\rho _2} \to {\rho _2}\left( {1 - \dfrac{{\Delta _{\varphi D}^{22} + 4\Delta _\square^{22}}}{{4{\Lambda ^2}}}} \right) - \dfrac{{\Delta _{\varphi D}^{12} + 4\Delta _\square^{12}}}{{4{\Lambda ^2}}}{\rho _1}\\
{\eta _1} \to {\eta _1}\left( {1 - \dfrac{{\Delta _{\varphi D}^{11}}}{{4{\Lambda ^2}}}} \right) - \dfrac{{\Delta _{\varphi D}^{12}}}{{4{\Lambda ^2}}}{\eta _2}\\
{\eta _2} \to {\eta _2}\left( {1 - \dfrac{{\Delta _{\varphi D}^{22}}}{{4{\Lambda ^2}}}} \right) - \dfrac{{\Delta _{\varphi D}^{12}}}{{4{\Lambda ^2}}}{\eta _1}\\
\phi _1^ +  \to \phi _1^ +  - \dfrac{{\Delta _{\varphi D}^ + }}{{4{\Lambda ^2}}}\phi _2^ + \\
\phi _2^ +  \to \phi _2^ +  - \dfrac{{\Delta _{\varphi D}^ + }}{{4{\Lambda ^2}}}\phi _1^ + \, .
\end{array}
\end{equation}

The operators with gauge and Higgs fields in Table~\ref{2higgs2bosons} lead to the shifts
\begin{eqnarray}
W_\mu ^ \pm  &\to& W_\mu ^ \pm \left( {1 + \frac{{{\Delta _{WW}}}}{{{\Lambda ^2}}}} \right)\\
{Z_\mu } &\to& {Z_\mu }\left( {1 + \frac{{{\Delta _{ZZ}}}}{{{\Lambda ^2}}}} \right) \nonumber \\
{A_\mu } &\to& {A_\mu }\left( {1 + \frac{{{\Delta _{AA}}}}{{{\Lambda ^2}}}} \right) + {Z_\mu }\frac{{{\Delta _{AZ}}}}{{{\Lambda ^2}}}\nonumber \\
G_\mu ^A &\to& G_\mu ^A\left( {1 + \frac{{{\Delta _{GG}}}}{{{\Lambda ^2}}}} \right) \nonumber
\end{eqnarray}
with
\begin{equation}
\renewcommand{\arraystretch}{1.7}
\begin{array}{l}
{\Delta _{GG}}= v_1^2\,C_{\varphi G}^{11} + v_2^2\,C_{\varphi G}^{22} \\ 
{\Delta _{WW}} = v_1^2\,C_{\varphi W}^{11} + v_2^2\,C_{\varphi W}^{22}\\
{\Delta _{ZZ}} = c_w^2\left( {v_1^2\,C_{\varphi W}^{11} + v_2^2\,C_{\varphi W}^{22}} \right) + s_w^2\left( {v_1^2\,C_{\varphi B}^{11} + v_2^2\,C_{\varphi B}^{22}} \right) + {c_w}{s_w}\left( {v_1^2\,C_{\varphi WB}^{11} + v_2^2\,C_{\varphi WB}^{22}} \right) \\
{\Delta _{AZ}} = 2{c_w}{s_w}\left[ {v_1^2\left( {C_{\varphi W}^{11} - C_{\varphi B}^{11}} \right) + v_2^2\left( {C_{\varphi W}^{22} - C_{\varphi B}^{22}} \right)} \right] + \left( {s_w^2 - c_w^2} \right)\left( {v_1^2C_{\varphi WB}^{11} + v_2^2\,C_{\varphi WB}^{22}} \right) \\
{\Delta _{AA}} = s_w^2\left( {v_1^2C_{\varphi W}^{11} + v_2^2C_{\varphi W}^{22}} \right) + c_w^2\left( {v_1^2\,C_{\varphi B}^{11} + v_2^2\,C_{\varphi B}^{22}} \right) - {c_w}{s_w}\left( {v_1^2C_{\varphi WB}^{11} + v_2^2C_{\varphi WB}^{22}} \right) \nonumber
\end{array}
\end{equation}
where $c_w$ and $s_w$ denote cosine and sine of the weak mixing angle, respectively.
These relations, together with the $\varphi D$ operators, affect the $W$ and $Z$ masses as
\begin{eqnarray}
M_Z^2 \!&=&\! \frac{1}{4}\left( {{g^2} + {{g'}^2}} \right)\left( {{v^2} + \frac{{4{v^2}\Delta _{ZZ}^{} + v_1^2\Delta _{\varphi D}^{11} + v_2^2\Delta _{\varphi D}^{22} + 2{v_1}{v_2}\Delta _{\varphi D}^{12}}}{{2{\Lambda ^2}}}} \right)\\
M_W^2 \!&=&\! \frac{1}{4}{g^2}\left( {{v^2} + \frac{{2{v^2}\Delta _{WW}^{} + {v_1}{v_2}\Delta _{\varphi D}^ + }}{{{\Lambda ^2}}}} \right)
\end{eqnarray}
where $v^2 \equiv v_1^2 + v_2^2$.

The shift of the Higgs fields applied to the dimension-four potential and the $\varphi^6$ operators modify the Higgs mass terms in the Lagrangian. Choosing again to eliminate $m_{11}^2$ and $m_{22}^2$ via the minimization conditions, we obtained
\begin{equation}
\begin{array}{{l}}
L_{{M_H}}^{\left( 4 \right)+\left( 6 \right)} = \dfrac{1}{2}{\left( {\begin{array}{*{20}{c}}
{{\eta _1}}\\
{{\eta _2}}
\end{array}} \right)^T}\left( {m_\eta ^2 + \Delta m_\eta ^2} \right)\left( {\begin{array}{*{20}{c}}
{{\eta _1}}\\
{{\eta _2}}
\end{array}} \right) + {\left( {\begin{array}{*{20}{c}}
{\phi _1^ - }\\
{\phi _2^ - }
\end{array}} \right)^T}\left( {m_{{\phi ^ \pm }}^2 + \Delta m_{{\phi ^ \pm }}^2} \right)\left( {\begin{array}{*{20}{c}}
{\phi _1^ + }\\
{\phi _2^ + }
\end{array}} \right)  \\
\qquad \qquad + \dfrac{1}{2}{\left( {\begin{array}{*{20}{c}}
{{\rho _1}}\\
{{\rho _2}}
\end{array}} \right)^T}\left( {m_\rho ^2 + \Delta m_\rho ^2} \right)\left( {\begin{array}{*{20}{c}}
{{\rho _1}}\\
{{\rho _2}}
\end{array}} \right) 
\end{array}
\end{equation}
with
\begin{equation}
\renewcommand{\arraystretch}{1.4}
\begin{array}{l}
\Delta m_\eta ^2 = \Delta m_{\varphi D\eta }^2 + \Delta m_{{\varphi ^6}\eta }^2 \\
\Delta m_\rho ^2 = \Delta m_{\varphi D\rho }^2 + \Delta m_{{\varphi ^6}\rho }^2 \\
\Delta m_{{\phi ^ \pm }}^2 = \Delta m_{\varphi D{\phi ^ \pm }}^2 + \Delta m_{{\varphi ^6}{\phi ^ \pm }}^2
\end{array}
\end{equation}
and
\begin{eqnarray}
\Delta m_{\varphi D{\phi ^ \pm }}^2 \!&=&\! \left[ {\left( {{\lambda _4} + {\lambda _5}} \right){v_1}{v_2} - 2m_{12}^2} \right]\frac{{\Delta _{\varphi {\rm{D}}}^ + }}{{4{\Lambda ^2}}}\left( {\begin{array}{*{20}{c}}
{ - 1}&{\dfrac{{{v^2}}}{{2{v_1}{v_2}}}}\\
{\dfrac{{{v^2}}}{{2{v_1}{v_2}}}}&{ - 1}
\end{array}} \right) \\
{\left( {\Delta m_{\varphi D\rho }^2} \right)_{11}} \!&=&\! \frac{{\left( {{v_1}\Delta _{\varphi {\rm{D}}}^{12} - {v_2}\Delta _{\varphi {\rm{D}}}^{11}} \right)m_{12}^2 - v_1^2\left[ {{v_1}\Delta _{\varphi {\rm{D}}}^{11}{\lambda _1} + {v_2}\Delta _{\varphi {\rm{D}}}^{12}\left( {{\lambda _3} + {\lambda _4} + {\lambda _5}} \right)} \right]}}{{2{\Lambda ^2}{v_1}}}\nonumber \\
{\left( {\Delta m_{\varphi D\rho }^2} \right)_{12}} \!&=&\! \frac{{\left[ { - {v^2}\Delta _{\varphi {\rm{D}}}^{12} + {v_1}{v_2}\left( {\Delta _{\varphi {\rm{D}}}^{22} + \Delta _{\varphi {\rm{D}}}^{11}} \right)} \right]m_{12}^2}}{{4{\Lambda ^2}{v_1}{v_2}}} \nonumber \\ &&+ \frac{{ {{v_1}{v_2}\left( { - \Delta _{\varphi {\rm{D}}}^{22} - \Delta _{\varphi {\rm{D}}}^{11}} \right)\left( {{\lambda _3} + {\lambda _4} + {\lambda _5}} \right) - \left( {v_1^2{\lambda _1} + v_2^2{\lambda _2}} \right)\Delta _{\varphi {\rm{D}}}^{12}} }}{{4{\Lambda ^2}}}\nonumber \\
{\left( {\Delta m_{\varphi D\rho }^2} \right)_{22}} \!&=&\! \dfrac{{\left( {{v_2}\Delta _{\varphi {\rm{D}}}^{12} - {v_1}\Delta _{\varphi {\rm{D}}}^{22}} \right)m_{12}^2 - v_2^2\left[ {{v_2}\Delta _{\varphi {\rm{D}}}^{22}{\lambda _2} + {v_1}\Delta _{\varphi {\rm{D}}}^{12}\left( {{\lambda _3} + {\lambda _4} + {\lambda _5}} \right)} \right]}}{{2{\Lambda ^2}{v_2}}}\nonumber \\
\Delta m_{\varphi D\eta }^2{\rm{ }} \!&=&\! \dfrac{{m_{12}^2 - {v_1}{v_2}{\lambda _5}}}{{2{\Lambda ^2}{v_1}{v_2}}}\left( {\begin{array}{*{20}{c}}
\!\!{{v_1}{v_2}\Delta _{\varphi {\rm{D}}}^{12} - v_2^2\Delta _{\varphi {\rm{D}}}^{11}}&\!\!{\dfrac{{{v_1}{v_2}\left( {\Delta _{\varphi {\rm{D}}}^{22} + \Delta _{\varphi {\rm{D}}}^{11}} \right) - {v^2}\Delta _{\varphi {\rm{D}}}^{12}}}{2}}\\
\!\!{\dfrac{{{v_1}{v_2}\left( {\Delta _{\varphi {\rm{D}}}^{22} + \Delta _{\varphi {\rm{D}}}^{11}} \right) - {v^2}\Delta _{\varphi {\rm{D}}}^{12}}}{2}}&\!\!{{v_1}{v_2}\Delta _{\varphi {\rm{D}}}^{12} - v_1^2\Delta _{\varphi {\rm{D}}}^{22}}
\end{array}} \right)\nonumber
\end{eqnarray}
as well as
\begin{eqnarray}
\Delta m_{{\varphi ^6}\eta }^2 \!&=&\! \Delta A\left( {\begin{array}{*{20}{c}}
{{{ - {v_2}} \mathord{\left/
 {\vphantom {{ - {v_2}} {{v_1}}}} \right.
 \kern-\nulldelimiterspace} {{v_1}}}}&1\nonumber \\
1&{{{ - {v_1}} \mathord{\left/
 {\vphantom {{ - {v_1}} {{v_2}}}} \right.
 \kern-\nulldelimiterspace} {{v_2}}}}
\end{array}} \right) \nonumber \\
\Delta m_{{\varphi ^6}{\phi ^ \pm }}^2 \!&=&\! \Delta {H^ \pm }\left( {\begin{array}{*{20}{c}}
{{{ - {v_2}} \mathord{\left/
 {\vphantom {{ - {v_2}} {{v_1}}}} \right.
 \kern-\nulldelimiterspace} {{v_1}}}}&1\\
1&{{{ - {v_1}} \mathord{\left/
 {\vphantom {{ - {v_1}} {{v_2}}}} \right.
 \kern-\nulldelimiterspace} {{v_2}}}}
\end{array}} \right)
\end{eqnarray}
with
\begin{eqnarray}
\Delta {H^ \pm } \!&=&\! \frac{{{v_1}{v_2}}}{{2{\Lambda ^2}}}\left[ {\left( {C_\varphi ^{\left( {1212} \right)1} + \frac{1}{2}C_\varphi ^{\left( {1221} \right)1}} \right)v_1^2} \right. + \left. {\left( {C_\varphi ^{\left( {1212} \right)2} + \frac{1}{2}C_\varphi ^{\left( {1221} \right)2}} \right)v_2^2} \right] \nonumber \\
\Delta A &=& \frac{{{v_1}{v_2}}}{{{\Lambda ^2}}}\left( {C_\varphi ^{\left( {1212} \right)1}v_1^2 + C_\varphi ^{\left( {1212} \right)2}v_2^2} \right)
\end{eqnarray}
and
\begin{eqnarray}
{\left( {\Delta m_{{\varphi ^6}\rho }^2} \right)_{11}} \!&=&\! \frac{{v_1^2}}{{{\Lambda ^2}}}\left( {3{\mkern 1mu} C_\varphi ^{111}v_1^2 + \left( {C_\varphi ^{112} + 2C_\varphi ^{(1212)1} + C_\varphi ^{(1221)1}} \right)v_2^2} \right)\nonumber \\
{\left( {\Delta m_{{\varphi ^6}\rho }^2} \right)_{12}} \!&=&\! \frac{{{v_1}{v_2}}}{{{\Lambda ^2}}}\left[ {\left( {C_\varphi ^{112} + 2C_\varphi ^{\left( {1212} \right)1} + C_\varphi ^{\left( {1221} \right)1}} \right)v_1^2} \right. + \left. {\left( {C_\varphi ^{122} + 2C_\varphi ^{\left( {1212} \right)2} + C_\varphi ^{\left( {1221} \right)2}} \right)v_2^2} \right]\nonumber \\
{\left( {\Delta m_{{\varphi ^6}\rho }^2} \right)_{22}} \!&=&\! \frac{{v_2^2}}{{{\Lambda ^2}}}\left[ {\left( {C_\varphi ^{122} + 2C_\varphi ^{\left( {1212} \right)2} + C_\varphi ^{\left( {1221} \right)2}} \right)v_1^2 + 3{\mkern 1mu} C_\varphi ^{222}v_2^2} \right]\, .
\end{eqnarray}

Interestingly, the mass matrices for the CP-odd and the charged Higgs bosons are not diagonalized anymore by the angle $\beta$ as in the case of the dimension-four potential. For this purpose, one needs two angles $\beta_\eta$ and $\beta_{\phi^\pm}$, which do not satisfy $\tan\beta_{\eta, \phi^\pm}=v_1/v_2$. This relation holds at dimension four but is broken by the $\varphi D$ operators. However, in the presence of $\square$ and $\varphi^6$ operators only, the CP-odd and the charged Higgs mass matrices would still be diagonalized by $\beta$ defined as $\tan\beta=v_1/v_2$. Also the angle $\alpha$ diagonalizing the CP-even mass matrix gets modified compared to the case of 2HDM without dimension-six operators. However, this effect can be accounted for by an appropriate redefinition of $\alpha$, which is anyway a free parameter in the 2HDM~\footnote{Nevertheless, one should keep in mind that in the limit of heavy Higgses $A^0,\,H^0,\,H^\pm$, $\alpha$ has to vanish as $\tan \beta \to \infty$ modulo corrections of order $v^2/m_H^2$. When dimension-six operators are included, this is still the case but additional (even smaller) corrections of order $v^2/\Lambda^2$ are present.}. The eigenvalues of the mass matrices also change, except those corresponding to the pseudo-Goldstone bosons, which are still zero at ${\cal O}(1/\Lambda^2)$, as we checked. Therefore, the effects on the eigenvalues can be absorbed into the definitions of $m_h$, $m_H$, $m_{H^\pm}$ and $m_A$.

\subsection{Yukawa sector}

After EW symmetry breaking, the fermion mass matrices in the presence of the dimension-six
operators of Table~\ref{2fermion3higgs} are given by
\begin{equation}
m^f = \frac{v_1Y_1^f}{\sqrt 2} + \frac{v_2Y_2^f}{\sqrt 2} + \frac{1}{2 \sqrt{2}\, \Lambda ^2}\left( v_1^3\,C_{f\varphi
}^{111} + v_1 v_2^2\,C_{f\varphi }^{122} + v_2^3\,C_{f\varphi }^{222} +
v_1^2 v_2\,C_{f\varphi }^{211} \right)\,,
\label{masses-yukawas}
\end{equation}
with $f=e,d,u$. Here $m^f$, $Y_{1,2}^f$ as well as the coefficients of the dimension-six operators are in general arbitrary $3\times 3$ matrices in flavor space. However, since the eigenvalues of $m^f$ are the physical fermion masses, by working in the basis where $m^f$ is diagonal, the rotations that map onto this basis get implicitly absorbed in the definitions of $Y_{1,2}^f$ and $C_{f\varphi }$.

Depending on which version of the four 2HDMs with natural flavor conservation we are interested in (or which one is assumed to be the limiting case of the 2HDM of type III~\footnote{Due to stringent flavor constraints (see Ref.~\cite{Crivellin:2013wna} for a recent analysis), a 2HDM with generic Yukawa couplings ({\it i.e.} of type III) should only slightly differ from one of the four 2HDMs with natural flavor conservation.}), one can choose to eliminate either $Y_1^f$ or $Y_2^f$ from the Yukawa Lagrangian. Afterwards, it is straightforward to calculate the couplings of $\phi_{1,2}^+$, $\eta_{1,2}$ and $\rho_{1,2}$ to fermions.

Taking as an example the lepton Yukawa couplings and eliminating $Y_1^e$, we find  
\bea
\renewcommand{\arraystretch}{2}
{\cal L}_{Y^e}^{(6)}&=&\dfrac{i}{{{v_1}}}\left( {{m^e} - \dfrac{{{v_2}}}{{2\sqrt 2 }}{\epsilon ^e}}
\right)\bar l {P_R}e{\eta _1}+\dfrac{i}{{2\sqrt 2 }}\,{\epsilon ^e}\bar l {P_R}e{\eta _2}\nonumber\\
&&+\dfrac{1}{{{v_1}}}\left( {\sqrt 2 {m^e} - \dfrac{{{v_2}}}{2}{\epsilon ^e}}
\right)\bar \nu {P_R}e\phi _1^ + \nonumber\\
&&+\dfrac{1}{2}\,{\epsilon ^e}\bar \nu {P_R}e\phi _2^+ +\left[ {\dfrac{{C_{e\varphi }^{111}v_1^2 + C_{e\varphi
}^{211}{{{v}}_1}{{{v}}_2}}}{{\sqrt 2 {\Lambda ^2}}} + \dfrac{1}{{{v_1}}}\left(
{{m^e} - \dfrac{{{{{v}}_2}{\epsilon^e}}}{{2\sqrt 2 }}} \right)} \right]\bar l
{P_R}e{\rho _1}\nonumber\\
&&+\left( {\dfrac{{C_{e\varphi }^{222}v_2^2 + C_{e\varphi
}^{122}{{{v}}_1}{{{v}}_2}}}{{\sqrt 2 {\Lambda ^2}}} +
\dfrac{{{\epsilon^e}}}{{2\sqrt 2 }}} \right)\bar l {P_R}e{\rho _2}
\label{Ylep}
\eea 
with
\begin{equation}
{\epsilon ^e} = 2Y_2^e + \frac{{C_{e\varphi }^{211}v_1^2 + C_{e\varphi
}^{222}v_2^2}}{{{\Lambda ^2}}}\,
\end{equation}
and $P_R$ denoting the right-handed projector. The outcome in the case that $Y_2^e$ is eliminated is obtained by simply interchanging $1\leftrightarrow 2$. In order to express these couplings in terms of the physical Higgs fields, the rotations by (redefined) $\alpha$, $\beta_\eta$ and $\beta_{\phi^\pm}$ have to be applied. The analogous of \eq{Ylep} for down-quarks has exactly the same structure. For up-type quarks all terms involving a charged or CP-odd Higgs switch sign and $\phi _{1,2}^+$ get replaced by $\phi _{1,2}^-$.

We can now point out an interesting effect arising in the 2HDM-EFT related to Higgs-pair production~\cite{Dawson:1998py,Djouadi:1999rca,Baglio:2012np,Frederix:2014hta}. In the SM-EFT there is only one operator giving rise to fermion-fermion-Higgs-Higgs interactions and this is directly correlated to fermion-fermion-Higgs couplings~\cite{Goertz:2014qta,Grober:2015cwa}. Therefore, its effect in the four-particle interaction is limited due to the constraints on the fermion-fermion-Higgs couplings. In our 2HDM-EFT three terms affect the fermion-fermion-Higgs couplings while the part of the Lagrangian relevant for Higgs pair production is
\begin{equation}
\frac{1}{{2\sqrt 2 {\Lambda ^2}}}\bar f\Big( {3{{C}}_{f \varphi} ^{111}{{{v}}_1}\rho _1^2 + {{C}}_{f \varphi} ^{122}{{{v}}_1}\rho _2^2 + {{C}}_{f \varphi} ^{122}{{{v}}_2}{\rho _1}{\rho _2} + {{C}}_{f \varphi} ^{211}{{{v}}_1}{\rho _1}{\rho _2} + {{C}}_{f \varphi} ^{211}{{{v}}_2}\rho _1^2 + 3{{C}}_{f \varphi} ^{222}{{{v}}_2}\rho _2^2} \Big)f\,.
\end{equation}
By comparing with \eq{Ylep}, it is clear that a cancellation between $C_{f \varphi}^{222}$ and $C_{f \varphi}^{122}$ could have quite different effects in fermion vertices with one or two Higgses. Therefore a suppressed modification to fermion-fermion-Higgs couplings accompanied by a sizable effect in pair production of SM-like Higgs bosons is possible in 2HDM-EFT.

\section{Conclusions and outlook}
\label{sec:conclusions}

In this article we extended the SM-EFT approach to NP to the case in which two Higgs doublets are dynamical degrees of freedom (2HDM-EFT). In this framework the effects of additional heavy particles ({\it e.g.} heavy SUSY partners in the
MSSM) are parameterized in terms of higher-dimension operators and their Wilson coefficients. Our analysis enables systematic studies of the role played by additional degrees of freedom beyond the 2HDM field content. We derived a complete set of independent gauge-invariant dimension-six operators with two Higgs doublets of the same hypercharge under the assumption of a $Z_2$ symmetry involving the Higgs field and the right-handed fermions. These operators modify the Higgs potential and the relation between the fermion masses and Higgs-fermion-fermion couplings. We performed the transition to the physical basis by re-diagonalizing the Higgs, gauge boson and fermion kinetic terms and mass matrices. We showed that the CP-odd and charged Higgs mass matrices are in general not diagonalized by one angle $\beta$ (defined by $\tan\beta=v_1/v_2$) but rather by two different angles. Finally, we derived the expressions for the Higgs-fermion-fermion couplings in the presence of dimension-six operators.

Even though a study of phenomenological applications of our framework is beyond the scope of this article, we pointed out one interesting example illustrating the differences between 2HDM-EFT and SM-EFT. In the latter, there is only one dimension-six operator giving rise to fermion-fermion-Higgs-Higgs interactions and its effect on pair production of SM Higgses is limited by constraints on Higgs-fermion-fermion couplings. In the 2HDM-EFT instead, there are three operators entering Higgs-fermion-fermion couplings, whose contributions can cancel each other. The same operators, as well as additional ones, also contribute to the couplings of Higgs pairs to fermions but with different prefactors. Higgs pair production is decoupled from single-Higgs-fermion interactions and could thus be sizable.

\vspace{1cm}
\noindent
{\bf Acknowledgements:  }

\noindent

We thank Michael Spira for useful discussions and for proofreading the article. A.C.~is supported by an Ambizione Fellowship of the Swiss National Science Foundation. M.P.~is supported by a Marie Curie Intra-European Fellowship of the European Community's 7th Framework Programme under contract number PIEF-GA-2013-622527.

\newpage

\label{Bibliography}
\bibliographystyle{JHEP}
 \footnotesize
\bibliography{BIB}

\providecommand{\href}[2]{#2}\begingroup\raggedright\begin{thebibliography}{10}

\bibitem{Lee:1973iz}
T.~D. Lee, {\it {A Theory of Spontaneous T Violation}},  {\em Phys. Rev.} {\bf
  D8} (1973) 1226--1239.

\bibitem{Gunion:1989we}
J.~F. Gunion, H.~E. Haber, G.~L. Kane, and S.~Dawson, {\it {The Higgs Hunter's
  Guide}},  {\em Front. Phys.} {\bf 80} (2000) 1--404.

\bibitem{Branco:2011iw}
G.~C. Branco, P.~M. Ferreira, L.~Lavoura, M.~N. Rebelo, M.~Sher, and J.~P.
  Silva, {\it {Theory and phenomenology of two-Higgs-doublet models}},  {\em
  Phys. Rept.} {\bf 516} (2012) 1--102,
  [\href{http://arxiv.org/abs/1106.0034}{{\tt arXiv:1106.0034}}].

\bibitem{Kim:1986ax}
J.~E. Kim, {\it {Light Pseudoscalars, Particle Physics and Cosmology}},  {\em
  Phys. Rept.} {\bf 150} (1987) 1--177.

\bibitem{Peccei:1977hh}
R.~D. Peccei and H.~R. Quinn, {\it {CP Conservation in the Presence of
  Instantons}},  {\em Phys. Rev. Lett.} {\bf 38} (1977) 1440--1443.

\bibitem{Trodden:1998ym}
M.~Trodden, {\it {Electroweak baryogenesis}},  {\em Rev. Mod. Phys.} {\bf 71}
  (1999) 1463--1500, [\href{http://arxiv.org/abs/hep-ph/9803479}{{\tt
  hep-ph/9803479}}].

\bibitem{Crivellin:2012ye}
A.~Crivellin, C.~Greub, and A.~Kokulu, {\it {Explaining $B\to D\tau\nu$, $B\to
  D^*\tau\nu$ and $B\to \tau\nu$ in a 2HDM of type III}},  {\em Phys. Rev.}
  {\bf D86} (2012) 054014, [\href{http://arxiv.org/abs/1206.2634}{{\tt
  arXiv:1206.2634}}].

\bibitem{Celis:2012dk}
A.~Celis, M.~Jung, X.-Q. Li, and A.~Pich, {\it {Sensitivity to charged scalars
  in $B\to D^{(*)}\tau\nu_\tau$ and $B\to\tau\nu_\tau$ decays}},  {\em JHEP}
  {\bf 01} (2013) 054, [\href{http://arxiv.org/abs/1210.8443}{{\tt
  arXiv:1210.8443}}].

\bibitem{Crivellin:2015hha}
A.~Crivellin, J.~Heeck, and P.~Stoffer, {\it {A perturbed lepton-specific
  two-Higgs-doublet model facing experimental hints for physics beyond the
  Standard Model}},  {\em Phys. Rev. Lett.} {\bf 116} (2016), no.~8 081801,
  [\href{http://arxiv.org/abs/1507.07567}{{\tt arXiv:1507.07567}}].

\bibitem{Fayet:1976cr}
P.~Fayet and S.~Ferrara, {\it {Supersymmetry}},  {\em Phys. Rept.} {\bf 32}
  (1977) 249--334.

\bibitem{Fayet:1976et}
P.~Fayet, {\it {Supersymmetry and Weak, Electromagnetic and Strong
  Interactions}},  {\em Phys. Lett.} {\bf B64} (1976) 159.

\bibitem{Haber:1984rc}
H.~E. Haber and G.~L. Kane, {\it {The Search for Supersymmetry: Probing Physics
  Beyond the Standard Model}},  {\em Phys. Rept.} {\bf 117} (1985) 75--263.

\bibitem{Carena:2013ytb}
M.~Carena, S.~Heinemeyer, O.~St\r{a}l, C.~E.~M. Wagner, and G.~Weiglein, {\it
  {MSSM Higgs Boson Searches at the LHC: Benchmark Scenarios after the
  Discovery of a Higgs-like Particle}},  {\em Eur. Phys. J.} {\bf C73} (2013),
  no.~9 2552, [\href{http://arxiv.org/abs/1302.7033}{{\tt arXiv:1302.7033}}].

\bibitem{PDG}
{\bf Particle Data Group} Collaboration, J.~Beringer et~al., {\it {Review of
  Particle Physics (RPP)}},  {\em Phys.Rev.} {\bf D86} (2012) 010001.

\bibitem{Misiak:2015xwa}
M.~Misiak et~al., {\it {Updated NNLO QCD predictions for the weak radiative
  B-meson decays}},  {\em Phys. Rev. Lett.} {\bf 114} (2015), no.~22 221801,
  [\href{http://arxiv.org/abs/1503.01789}{{\tt arXiv:1503.01789}}].

\bibitem{Crivellin:2015mga}
A.~Crivellin, G.~D'Ambrosio, and J.~Heeck, {\it {Explaining
  $h\to\mu^\pm\tau^\mp$, $B\to K^* \mu^+\mu^-$ and $B\to K \mu^+\mu^-/B\to K
  e^+e^-$ in a two-Higgs-doublet model with gauged $L_\mu-L_\tau$}},  {\em
  Phys. Rev. Lett.} {\bf 114} (2015) 151801,
  [\href{http://arxiv.org/abs/1501.00993}{{\tt arXiv:1501.00993}}].

\bibitem{Crivellin:2015lwa}
A.~Crivellin, G.~D'Ambrosio, and J.~Heeck, {\it {Addressing the LHC flavor
  anomalies with horizontal gauge symmetries}},  {\em Phys. Rev.} {\bf D91}
  (2015), no.~7 075006, [\href{http://arxiv.org/abs/1503.03477}{{\tt
  arXiv:1503.03477}}].

\bibitem{Weinberg:1979sa}
S.~Weinberg, {\it {Baryon and Lepton Nonconserving Processes}},  {\em Phys.
  Rev. Lett.} {\bf 43} (1979) 1566--1570.

\bibitem{Buchmuller:1985jz}
W.~Buchmuller and D.~Wyler, {\it {Effective Lagrangian Analysis of New
  Interactions and Flavor Conservation}},  {\em Nucl.Phys.} {\bf B268} (1986)
  621.

\bibitem{Burges:1983zg}
C.~J.~C. Burges and H.~J. Schnitzer, {\it {Virtual Effects of Excited Quarks as
  Probes of a Possible New Hadronic Mass Scale}},  {\em Nucl. Phys.} {\bf B228}
  (1983) 464--500.

\bibitem{Leung:1984ni}
C.~N. Leung, S.~T. Love, and S.~Rao, {\it {Low-Energy Manifestations of a New
  Interaction Scale: Operator Analysis}},  {\em Z. Phys.} {\bf C31} (1986) 433.

\bibitem{Hagiwara:1993qt}
K.~Hagiwara, R.~Szalapski, and D.~Zeppenfeld, {\it {Anomalous Higgs boson
  production and decay}},  {\em Phys.Lett.} {\bf B318} (1993) 155--162,
  [\href{http://arxiv.org/abs/hep-ph/9308347}{{\tt hep-ph/9308347}}].

\bibitem{GIMR}
B.~Grzadkowski, M.~Iskrzynski, M.~Misiak, and J.~Rosiek, {\it {Dimension-Six
  Terms in the Standard Model Lagrangian}},  {\em JHEP} {\bf 1010} (2010) 085,
  [\href{http://arxiv.org/abs/1008.4884}{{\tt arXiv:1008.4884}}].

\bibitem{Degrande:2012wf}
C.~Degrande, N.~Greiner, W.~Kilian, O.~Mattelaer, H.~Mebane, et~al., {\it
  {Effective Field Theory: A Modern Approach to Anomalous Couplings}},  {\em
  Annals Phys.} {\bf 335} (2013) 21--32,
  [\href{http://arxiv.org/abs/1205.4231}{{\tt arXiv:1205.4231}}].

\bibitem{Trott:2014dma}
M.~Trott, {\it {On the consistent use of Constructed Observables}},  {\em JHEP}
  {\bf 1502} (2015) 046, [\href{http://arxiv.org/abs/1409.7605}{{\tt
  arXiv:1409.7605}}].

\bibitem{Masso:2014xra}
E.~Masso, {\it {An Effective Guide to Beyond the Standard Model Physics}},
  {\em JHEP} {\bf 1410} (2014) 128, [\href{http://arxiv.org/abs/1406.6376}{{\tt
  arXiv:1406.6376}}].

\bibitem{Henning:2014wua}
B.~Henning, X.~Lu, and H.~Murayama, {\it {How to use the Standard Model
  effective field theory}},  {\em JHEP} {\bf 01} (2016) 023,
  [\href{http://arxiv.org/abs/1412.1837}{{\tt arXiv:1412.1837}}].

\bibitem{Contino:2016jqw}
R.~Contino, A.~Falkowski, F.~Goertz, C.~Grojean, and F.~Riva, {\it {On the
  Validity of the Effective Field Theory Approach to SM Precision Tests}},
  \href{http://arxiv.org/abs/1604.06444}{{\tt arXiv:1604.06444}}.

\bibitem{Giudice:2007fh}
G.~F. Giudice, C.~Grojean, A.~Pomarol, and R.~Rattazzi, {\it {The
  Strongly-Interacting Light Higgs}},  {\em JHEP} {\bf 06} (2007) 045,
  [\href{http://arxiv.org/abs/hep-ph/0703164}{{\tt hep-ph/0703164}}].

\bibitem{Contino:2013kra}
R.~Contino, M.~Ghezzi, C.~Grojean, M.~M{\"u}hlleitner, and M.~Spira, {\it
  {Effective Lagrangian for a light Higgs-like scalar}},  {\em JHEP} {\bf 1307}
  (2013) 035, [\href{http://arxiv.org/abs/1303.3876}{{\tt arXiv:1303.3876}}].

\bibitem{Alloul:2013naa}
A.~Alloul, B.~Fuks, and V.~Sanz, {\it {Phenomenology of the Higgs Effective
  Lagrangian via FEYNRULES}},  {\em JHEP} {\bf 1404} (2014) 110,
  [\href{http://arxiv.org/abs/1310.5150}{{\tt arXiv:1310.5150}}].

\bibitem{Contino:2014aaa}
R.~Contino, M.~Ghezzi, C.~Grojean, M.~M{\"u}hlleitner, and M.~Spira, {\it
  {eHDECAY: an Implementation of the Higgs Effective Lagrangian into HDECAY}},
  {\em Comput. Phys. Commun.} {\bf 185} (2014) 3412--3423,
  [\href{http://arxiv.org/abs/1403.3381}{{\tt arXiv:1403.3381}}].

\bibitem{Chen:2013kfa}
C.-Y. Chen, S.~Dawson, and C.~Zhang, {\it {Electroweak Effective Operators and
  Higgs Physics}},  {\em Phys.Rev.} {\bf D89} (2014) 015016,
  [\href{http://arxiv.org/abs/1311.3107}{{\tt arXiv:1311.3107}}].

\bibitem{Englert:2014uua}
C.~Englert, A.~Freitas, M.~M. M{\"u}hlleitner, T.~Plehn, M.~Rauch, M.~Spira,
  and K.~Walz, {\it {Precision Measurements of Higgs Couplings: Implications
  for New Physics Scales}},  {\em J. Phys.} {\bf G41} (2014) 113001,
  [\href{http://arxiv.org/abs/1403.7191}{{\tt arXiv:1403.7191}}].

\bibitem{Biekoetter:2014jwa}
A.~Biekotter, A.~Knochel, M.~Kraemer, D.~Liu, and F.~Riva, {\it {Vices and
  virtues of Higgs effective field theories at large energy}},  {\em Phys.Rev.}
  {\bf D91} (2015) 055029, [\href{http://arxiv.org/abs/1406.7320}{{\tt
  arXiv:1406.7320}}].

\bibitem{Bonnet:2011yx}
F.~Bonnet, M.~B. Gavela, T.~Ota, and W.~Winter, {\it {Anomalous Higgs couplings
  at the LHC, and their theoretical interpretation}},  {\em Phys. Rev.} {\bf
  D85} (2012) 035016, [\href{http://arxiv.org/abs/1105.5140}{{\tt
  arXiv:1105.5140}}].

\bibitem{Bonnet:2012nm}
F.~Bonnet, T.~Ota, M.~Rauch, and W.~Winter, {\it {Interpretation of precision
  tests in the Higgs sector in terms of physics beyond the Standard Model}},
  {\em Phys. Rev.} {\bf D86} (2012) 093014,
  [\href{http://arxiv.org/abs/1207.4599}{{\tt arXiv:1207.4599}}].

\bibitem{Crivellin:2013hpa}
A.~Crivellin, S.~Najjari, and J.~Rosiek, {\it {Lepton Flavor Violation in the
  Standard Model with general Dimension-Six Operators}},  {\em JHEP} {\bf 1404}
  (2014) 167, [\href{http://arxiv.org/abs/1312.0634}{{\tt arXiv:1312.0634}}].

\bibitem{Crivellin:2014cta}
A.~Crivellin, M.~Hoferichter, and M.~Procura, {\it {Improved predictions for
  $\mu\to e$ conversion in nuclei and Higgs-induced lepton flavor violation}},
  {\em Phys.Rev.} {\bf D89} (2014) 093024,
  [\href{http://arxiv.org/abs/1404.7134}{{\tt arXiv:1404.7134}}].

\bibitem{Crivellin:2014zpa}
A.~Crivellin and S.~Pokorski, {\it {Can the differences in the determinations
  of $V_{ub}$ and $V_{cb}$ be explained by New Physics?}},  {\em Phys. Rev.
  Lett.} {\bf 114} (2015), no.~1 011802,
  [\href{http://arxiv.org/abs/1407.1320}{{\tt arXiv:1407.1320}}].

\bibitem{Pruna:2014asa}
G.~M. Pruna and A.~Signer, {\it {The $\mu\to e\gamma$ decay in a systematic
  effective field theory approach with dimension 6 operators}},  {\em JHEP}
  {\bf 1410} (2014) 14, [\href{http://arxiv.org/abs/1408.3565}{{\tt
  arXiv:1408.3565}}].

\bibitem{Bhattacharya:2014wla}
B.~Bhattacharya, A.~Datta, D.~London, and S.~Shivashankara, {\it {Simultaneous
  Explanation of the $R_K$ and $R(D^{(*)})$ Puzzles}},  {\em Phys.Lett.} {\bf
  B742} (2015) 370--374, [\href{http://arxiv.org/abs/1412.7164}{{\tt
  arXiv:1412.7164}}].

\bibitem{Alonso:2014csa}
R.~Alonso, B.~Grinstein, and J.~{Martin Camalich}, {\it {$SU(2)\times U(1)$
  gauge invariance and the shape of new physics in rare $B$ decays}},  {\em
  Phys.Rev.Lett.} {\bf 113} (2014) 241802,
  [\href{http://arxiv.org/abs/1407.7044}{{\tt arXiv:1407.7044}}].

\bibitem{Buras:2014fpa}
A.~J. Buras, J.~Girrbach-Noe, C.~Niehoff, and D.~M. Straub, {\it {$ B\to
  {K}^{\left(*\right)}\nu \overline{\nu} $ decays in the Standard Model and
  beyond}},  {\em JHEP} {\bf 1502} (2015) 184,
  [\href{http://arxiv.org/abs/1409.4557}{{\tt arXiv:1409.4557}}].

\bibitem{Alonso:2015sja}
R.~Alonso, B.~Grinstein, and J.~Martin~Camalich, {\it {Lepton universality
  violation and lepton flavor conservation in $B$-meson decays}},  {\em JHEP}
  {\bf 10} (2015) 184, [\href{http://arxiv.org/abs/1505.05164}{{\tt
  arXiv:1505.05164}}].

\bibitem{Calibbi:2015kma}
L.~Calibbi, A.~Crivellin, and T.~Ota, {\it {Effective field theory approach to
  $b\to s\ell\ell^{(\prime)}$, $B\to K^{(*)}\nu\bar{\nu}$ and $B\to
  D^{(*)}\tau\nu$ with third generation couplings}},  {\em Phys. Rev. Lett.}
  {\bf 115} (2015) 181801, [\href{http://arxiv.org/abs/1506.02661}{{\tt
  arXiv:1506.02661}}].

\bibitem{Aebischer:2015fzz}
J.~Aebischer, A.~Crivellin, M.~Fael, and C.~Greub, {\it {Matching of gauge
  invariant dimension-six operators for $b\to s$ and $b\to c$ transitions}},
  {\em JHEP} {\bf 05} (2016) 037, [\href{http://arxiv.org/abs/1512.02830}{{\tt
  arXiv:1512.02830}}].

\bibitem{Berthier:2015oma}
L.~Berthier and M.~Trott, {\it {Towards consistent Electroweak Precision Data
  constraints in the SMEFT}},  {\em JHEP} {\bf 05} (2015) 024,
  [\href{http://arxiv.org/abs/1502.02570}{{\tt arXiv:1502.02570}}].

\bibitem{Ellis:2014dva}
J.~Ellis, V.~Sanz, and T.~You, {\it {Complete Higgs Sector Constraints on
  Dimension-6 Operators}},  {\em JHEP} {\bf 1407} (2014) 036,
  [\href{http://arxiv.org/abs/1404.3667}{{\tt arXiv:1404.3667}}].

\bibitem{Falkowski:2014tna}
A.~Falkowski and F.~Riva, {\it {Model-independent precision constraints on
  dimension-6 operators}},  {\em JHEP} {\bf 1502} (2015) 039,
  [\href{http://arxiv.org/abs/1411.0669}{{\tt arXiv:1411.0669}}].

\bibitem{Low:2009di}
I.~Low, R.~Rattazzi, and A.~Vichi, {\it {Theoretical Constraints on the Higgs
  Effective Couplings}},  {\em JHEP} {\bf 1004} (2010) 126,
  [\href{http://arxiv.org/abs/0907.5413}{{\tt arXiv:0907.5413}}].

\bibitem{Pomarol:2013zra}
A.~Pomarol and F.~Riva, {\it {Towards the Ultimate SM Fit to Close in on Higgs
  Physics}},  {\em JHEP} {\bf 1401} (2014) 151,
  [\href{http://arxiv.org/abs/1308.2803}{{\tt arXiv:1308.2803}}].

\bibitem{Berthier:2015gja}
L.~Berthier and M.~Trott, {\it {Consistent constraints on the Standard Model
  Effective Field Theory}},  {\em JHEP} {\bf 02} (2016) 069,
  [\href{http://arxiv.org/abs/1508.05060}{{\tt arXiv:1508.05060}}].

\bibitem{Englert:2015hrx}
C.~Englert, R.~Kogler, H.~Schulz, and M.~Spannowsky, {\it {Higgs coupling
  measurements at the LHC}},  \href{http://arxiv.org/abs/1511.05170}{{\tt
  arXiv:1511.05170}}.

\bibitem{Cirigliano:2016nyn}
V.~Cirigliano, W.~Dekens, J.~de~Vries, and E.~Mereghetti, {\it {Constraining
  the top-Higgs sector of the Standard Model Effective Field Theory}},
  \href{http://arxiv.org/abs/1605.04311}{{\tt arXiv:1605.04311}}.

\bibitem{Grojean:2013kd}
C.~Grojean, E.~E. Jenkins, A.~V. Manohar, and M.~Trott, {\it {Renormalization
  Group Scaling of Higgs Operators and $\Gamma(h \to \gamma \gamma)$}},  {\em
  JHEP} {\bf 04} (2013) 016, [\href{http://arxiv.org/abs/1301.2588}{{\tt
  arXiv:1301.2588}}].

\bibitem{Jenkins:2013zja}
E.~E. Jenkins, A.~V. Manohar, and M.~Trott, {\it {Renormalization Group
  Evolution of the Standard Model Dimension Six Operators I: Formalism and
  lambda Dependence}},  {\em JHEP} {\bf 1310} (2013) 087,
  [\href{http://arxiv.org/abs/1308.2627}{{\tt arXiv:1308.2627}}].

\bibitem{Jenkins:2013wua}
E.~E. Jenkins, A.~V. Manohar, and M.~Trott, {\it {Renormalization Group
  Evolution of the Standard Model Dimension Six Operators II: Yukawa
  Dependence}},  {\em JHEP} {\bf 01} (2014) 035,
  [\href{http://arxiv.org/abs/1310.4838}{{\tt arXiv:1310.4838}}].

\bibitem{Alonso:2013hga}
R.~Alonso, E.~E. Jenkins, A.~V. Manohar, and M.~Trott, {\it {Renormalization
  Group Evolution of the Standard Model Dimension Six Operators III: Gauge
  Coupling Dependence and Phenomenology}},  {\em JHEP} {\bf 04} (2014) 159,
  [\href{http://arxiv.org/abs/1312.2014}{{\tt arXiv:1312.2014}}].

\bibitem{Grober:2015cwa}
R.~Gr{\"o}ber, M.~M{\"u}hlleitner, M.~Spira, and J.~Streicher, {\it {NLO QCD
  Corrections to Higgs Pair Production including Dimension-6 Operators}},  {\em
  JHEP} {\bf 09} (2015) 092, [\href{http://arxiv.org/abs/1504.06577}{{\tt
  arXiv:1504.06577}}].

\bibitem{Elias-Miro:2013mua}
J.~Elias-Mir\'o, J.~Espinosa, E.~Masso, and A.~Pomarol, {\it {Higgs windows to
  new physics through d=6 operators: constraints and one-loop anomalous
  dimensions}},  {\em JHEP} {\bf 1311} (2013) 066,
  [\href{http://arxiv.org/abs/1308.1879}{{\tt arXiv:1308.1879}}].

\bibitem{Elias-Miro:2013gya}
J.~Elias-Mir\'o, J.~Espinosa, E.~Masso, and A.~Pomarol, {\it {Renormalization
  of dimension-six operators relevant for the Higgs decays $h\rightarrow
  \gamma\gamma,\gamma Z$}},  {\em JHEP} {\bf 1308} (2013) 033,
  [\href{http://arxiv.org/abs/1302.5661}{{\tt arXiv:1302.5661}}].

\bibitem{Passarino:2012cb}
G.~Passarino, {\it {NLO Inspired Effective Lagrangians for Higgs Physics}},
  {\em Nucl.Phys.} {\bf B868} (2013) 416--458,
  [\href{http://arxiv.org/abs/1209.5538}{{\tt arXiv:1209.5538}}].

\bibitem{Ghezzi:2015vva}
M.~Ghezzi, R.~Gomez-Ambrosio, G.~Passarino, and S.~Uccirati, {\it {NLO Higgs
  effective field theory and $\kappa$-framework}},  {\em JHEP} {\bf 07} (2015)
  175, [\href{http://arxiv.org/abs/1505.03706}{{\tt arXiv:1505.03706}}].

\bibitem{Hartmann:2015oia}
C.~Hartmann and M.~Trott, {\it {On one-loop corrections in the standard model
  effective field theory; the $\Gamma(h \rightarrow \gamma \, \gamma)$ case}},
  {\em JHEP} {\bf 07} (2015) 151, [\href{http://arxiv.org/abs/1505.02646}{{\tt
  arXiv:1505.02646}}].

\bibitem{Hartmann:2015aia}
C.~Hartmann and M.~Trott, {\it {Higgs Decay to Two Photons at One Loop in the
  Standard Model Effective Field Theory}},  {\em Phys. Rev. Lett.} {\bf 115}
  (2015), no.~19 191801, [\href{http://arxiv.org/abs/1507.03568}{{\tt
  arXiv:1507.03568}}].

\bibitem{Gauld:2015lmb}
R.~Gauld, B.~D. Pecjak, and D.~J. Scott, {\it {One-loop corrections to $h\to
  b\bar b$ and $h\to \tau\bar \tau$ decays in the Standard Model Dimension-6
  EFT: four-fermion operators and the large-$m_t$ limit}},  {\em JHEP} {\bf 05}
  (2016) 080, [\href{http://arxiv.org/abs/1512.02508}{{\tt arXiv:1512.02508}}].

\bibitem{Aaboud:2016tru}
{\bf ATLAS} Collaboration, M.~Aaboud et~al., {\it {Search for resonances in
  diphoton events at $\sqrt{s}$=13 TeV with the ATLAS detector}},
  \href{http://arxiv.org/abs/1606.03833}{{\tt arXiv:1606.03833}}.

\bibitem{Khachatryan:2016hje}
{\bf CMS} Collaboration, V.~Khachatryan et~al., {\it {Search for resonant
  production of high-mass photon pairs in proton-proton collisions at sqrt(s) =
  8 and 13 TeV}},  \href{http://arxiv.org/abs/1606.04093}{{\tt
  arXiv:1606.04093}}.

\bibitem{Altmannshofer:2015xfo}
W.~Altmannshofer, J.~Galloway, S.~Gori, A.~L. Kagan, A.~Martin, and J.~Zupan,
  {\it {750 GeV diphoton excess}},  {\em Phys. Rev.} {\bf D93} (2016), no.~9
  095015, [\href{http://arxiv.org/abs/1512.07616}{{\tt arXiv:1512.07616}}].

\bibitem{Gupta:2015zzs}
R.~S. Gupta, S.~Jager, Y.~Kats, G.~Perez, and E.~Stamou, {\it {Interpreting a
  750 GeV Diphoton Resonance}},  \href{http://arxiv.org/abs/1512.05332}{{\tt
  arXiv:1512.05332}}.

\bibitem{Bizot:2015qqo}
N.~Bizot, S.~Davidson, M.~Frigerio, and J.~L. Kneur, {\it {Two Higgs doublets
  to explain the excesses $pp\rightarrow \gamma\gamma(750\ {\rm GeV})$ and $h
  \to \tau^\pm \mu^\mp$}},  {\em JHEP} {\bf 03} (2016) 073,
  [\href{http://arxiv.org/abs/1512.08508}{{\tt arXiv:1512.08508}}].

\bibitem{Han:2016bvl}
X.-F. Han, L.~Wang, and J.~M. Yang, {\it {An extension of two-Higgs-doublet
  model and the excesses of 750 GeV diphoton, muon g-2 and $h\to\mu\tau$}},
  {\em Phys. Lett.} {\bf B757} (2016) 537--547,
  [\href{http://arxiv.org/abs/1601.04954}{{\tt arXiv:1601.04954}}].

\bibitem{Angelescu:2015uiz}
A.~Angelescu, A.~Djouadi, and G.~Moreau, {\it {Scenarii for interpretations of
  the LHC diphoton excess: two Higgs doublets and vector-like quarks and
  leptons}},  {\em Phys. Lett.} {\bf B756} (2016) 126--132,
  [\href{http://arxiv.org/abs/1512.04921}{{\tt arXiv:1512.04921}}].

\bibitem{DiChiara:2015vdm}
S.~Di~Chiara, L.~Marzola, and M.~Raidal, {\it {First interpretation of the 750
  GeV diphoton resonance at the LHC}},  {\em Phys. Rev.} {\bf D93} (2016),
  no.~9 095018, [\href{http://arxiv.org/abs/1512.04939}{{\tt
  arXiv:1512.04939}}].

\bibitem{Low:2015qep}
M.~Low, A.~Tesi, and L.-T. Wang, {\it {A pseudoscalar decaying to photon pairs
  in the early LHC Run 2 data}},  {\em JHEP} {\bf 03} (2016) 108,
  [\href{http://arxiv.org/abs/1512.05328}{{\tt arXiv:1512.05328}}].

\bibitem{Moretti:2015pbj}
S.~Moretti and K.~Yagyu, {\it {750 GeV diphoton excess and its explanation in
  two-Higgs-doublet models with a real inert scalar multiplet}},  {\em Phys.
  Rev.} {\bf D93} (2016), no.~5 055043,
  [\href{http://arxiv.org/abs/1512.07462}{{\tt arXiv:1512.07462}}].

\bibitem{Bertuzzo:2016fmv}
E.~Bertuzzo, P.~A.~N. Machado, and M.~Taoso, {\it {Di-Photon excess in the
  2HDM: hasting towards the instability and the non-perturbative regime}},
  \href{http://arxiv.org/abs/1601.07508}{{\tt arXiv:1601.07508}}.

\bibitem{Hernandez:2016rbi}
A.~E. Carcamo~Hernandez, I.~de~Medeiros~Varzielas, and E.~Schumacher, {\it {The
  $750\,\text{GeV}$ diphoton resonance in the light of a 2HDM with $S_3$
  flavour symmetry}},  \href{http://arxiv.org/abs/1601.00661}{{\tt
  arXiv:1601.00661}}.

\bibitem{Perez:1995dc}
M.~A. Perez, J.~J. Toscano, and J.~Wudka, {\it {Two photon processes and
  effective Lagrangians with an extended scalar sector}},  {\em Phys. Rev.}
  {\bf D52} (1995) 494--504, [\href{http://arxiv.org/abs/hep-ph/9506457}{{\tt
  hep-ph/9506457}}].

\bibitem{Gunion:2002zf}
J.~F. Gunion and H.~E. Haber, {\it {The CP conserving two Higgs doublet model:
  The Approach to the decoupling limit}},  {\em Phys. Rev.} {\bf D67} (2003)
  075019, [\href{http://arxiv.org/abs/hep-ph/0207010}{{\tt hep-ph/0207010}}].

\bibitem{Bjorken:1977vt}
J.~D. Bjorken and S.~Weinberg, {\it {A Mechanism for Nonconservation of Muon
  Number}},  {\em Phys. Rev. Lett.} {\bf 38} (1977) 622.

\bibitem{McWilliams:1980kj}
B.~McWilliams and L.-F. Li, {\it {Virtual Effects of Higgs Particles}},  {\em
  Nucl. Phys.} {\bf B179} (1981) 62--84.

\bibitem{Cheng:1987rs}
T.~P. Cheng and M.~Sher, {\it {Mass Matrix Ansatz and Flavor Nonconservation in
  Models with Multiple Higgs Doublets}},  {\em Phys. Rev.} {\bf D35} (1987)
  3484.

\bibitem{Crivellin:2013wna}
A.~Crivellin, A.~Kokulu, and C.~Greub, {\it {Flavor-phenomenology of
  two-Higgs-doublet models with generic Yukawa structure}},  {\em Phys. Rev.}
  {\bf D87} (2013), no.~9 094031, [\href{http://arxiv.org/abs/1303.5877}{{\tt
  arXiv:1303.5877}}].

\bibitem{Dawson:1998py}
S.~Dawson, S.~Dittmaier, and M.~Spira, {\it {Neutral Higgs boson pair
  production at hadron colliders: QCD corrections}},  {\em Phys. Rev.} {\bf
  D58} (1998) 115012, [\href{http://arxiv.org/abs/hep-ph/9805244}{{\tt
  hep-ph/9805244}}].

\bibitem{Djouadi:1999rca}
A.~Djouadi, W.~Kilian, M.~M{\"u}hlleitner, and P.~M. Zerwas, {\it {Production
  of neutral Higgs boson pairs at LHC}},  {\em Eur. Phys. J.} {\bf C10} (1999)
  45--49, [\href{http://arxiv.org/abs/hep-ph/9904287}{{\tt hep-ph/9904287}}].

\bibitem{Baglio:2012np}
J.~Baglio, A.~Djouadi, R.~Gr{\"o}ber, M.~M. M{\"u}hlleitner, J.~Quevillon, and
  M.~Spira, {\it {The measurement of the Higgs self-coupling at the LHC:
  theoretical status}},  {\em JHEP} {\bf 04} (2013) 151,
  [\href{http://arxiv.org/abs/1212.5581}{{\tt arXiv:1212.5581}}].

\bibitem{Frederix:2014hta}
R.~Frederix, S.~Frixione, V.~Hirschi, F.~Maltoni, O.~Mattelaer, P.~Torrielli,
  E.~Vryonidou, and M.~Zaro, {\it {Higgs pair production at the LHC with NLO
  and parton-shower effects}},  {\em Phys. Lett.} {\bf B732} (2014) 142--149,
  [\href{http://arxiv.org/abs/1401.7340}{{\tt arXiv:1401.7340}}].

\bibitem{Goertz:2014qta}
F.~Goertz, A.~Papaefstathiou, L.~L. Yang, and J.~Zurita, {\it {Higgs boson pair
  production in the D=6 extension of the SM}},  {\em JHEP} {\bf 04} (2015) 167,
  [\href{http://arxiv.org/abs/1410.3471}{{\tt arXiv:1410.3471}}].

\end{thebibliography}\endgroup
\end{document}